\documentclass[fleqn,usenatbib]{mnras}

\usepackage{newtxtext,newtxmath}

\usepackage[T1]{fontenc}

\DeclareRobustCommand{\VAN}[3]{#2}
\let\VANthebibliography\thebibliography
\def\thebibliography{\DeclareRobustCommand{\VAN}[3]{##3}\VANthebibliography}

\usepackage{graphicx}	
\usepackage{amsmath}	
\usepackage{lipsum}
\usepackage{subcaption}
\usepackage{lscape} 
\usepackage{placeins}   
\usepackage{hyperref}
\usepackage{cleveref}
\usepackage{comment}   
\usepackage{natbib}
\usepackage{adjustbox} 
\usepackage{gensymb}
\usepackage{color}
\usepackage{ulem} 

\crefname{figure}{Figure}{Figure}    
\crefname{table}{Table.}{Tables.}  
\crefname{equation}{Eq.}{Eqs.}   
\crefname{subfigure}{Figure}{Figure} 
\newcommand{\D}{\mathcal{D}} 

\title[Turbulence profiling with MTM]{Atmospheric turbulence profiling with the Multistar Turbulence Monitor}

\author[Huang et al.]
{
  Weisen Huang,$^{1,2}$ 
  Bin Ma,$^{1}$\thanks{E-mail: mabin3@mail.sysu.edu.cn (B. Ma)}
  Tengfei Song,$^{2}$\thanks{E-mail: stf@ynao.ac.cn (T. Song)}
  Paul Hickson,$^{3}$
  Zhaohui Shang,$^{4}$
  Xuefei Zhang,$^{2}$
  \newauthor
  Mingyu Zhao$^{2}$
  and Qing Zhou$^{5}$
  \\
  $^{1}$School of Physics and Astronomy, Sun Yat-Sen University, Zhuhai 519082, China\\
  $^{2}$Yunnan Observatory, Chinese Academy of Sciences (CAS), Kunming 650011, Yunnan, China\\
  $^{3}$Department of Physics and Astronomy, University of British Columbia, 6224 Agricultural Road, Vancouver, BC V6T 1Z1, Canada\\
  $^{4}$National Astronomical Observatories, Chinese Academy of Sciences, 20A Datun Road, Chaoyang District, Beijing 100101, China\\
  $^{5}$School of Physical Science and Technology, Southwest Jiaotong University, Chengdu 611756, China
}

\date{Accepted XXX. Received YYY; in original form ZZZ}

\pubyear{\the\year{}}

\begin{document}
\label{firstpage}
\pagerange{\pageref{firstpage}--\pageref{lastpage}}
\maketitle

\begin{abstract}
Accurate characterization of atmospheric optical turbulence is essential for evaluating astronomical sites and optimizing adaptive optics systems. The Multistar Turbulence Monitor (MTM) infers the vertical distribution of the refractive-index structure constant $C_n^2(z)$ from differential image motion measured between multiple stellar pairs in short-exposure frames. We present a comprehensive investigation of the MTM method, combining theoretical analysis, instrument-performance assessment, numerical simulations, and on-sky observations obtained at the Daocheng Astronomical Site. Simulations based on a standard Hufnagel–Valley turbulence model demonstrate that the inversion pipeline robustly recovers both the integrated seeing and the vertical turbulence profile under realistic centroiding noise and varying pixel scales. The Markov Chain Monte Carlo (MCMC) inversion achieves stable results with thirteen discrete height nodes and provides reliable uncertainties. Three nights of MTM measurements at the Daocheng Astronomical Site show that MTM-derived seeing closely tracks simultaneous Differential Image Motion Monitor (DIMM) results, accurately reproducing both short-term fluctuations and nightly averages. These results confirm that MTM provides a simple, portable, and versatile solution for atmospheric turbulence profiling and routine seeing monitoring.
\end{abstract}




\section{Introduction}
 
Atmospheric turbulence significantly affects light propagation, severely degrading the imaging quality of ground-based telescopes. By inducing random fluctuations in the atmospheric refractive index, turbulence causes image blurring and motion. This significantly restricts the angular resolution of ground-based telescopes, necessitating Adaptive Optics (AO) to compensate for these distortions \citep{Roddier1981, Beckers1993a}. The measurement of turbulent parameters is paramount for optimizing astronomical observations and selecting astronomical sites. Consequently, a variety of techniques have been developed for this purpose \citep{Deng2023}. Likewise, accurate estimation of these parameters is fundamental for the dimensioning of AO systems and their performance predictions. In the era of next-generation extremely large telescopes, precise atmospheric characterization is becoming increasingly critical. Major international projects, such as the European Extremely Large Telescope \citep[E-ELT;][]{Vernin2011}, the Thirty Meter Telescope \citep[TMT;][]{Schock2009}, and the Giant Magellan Telescope \citep[GMT;][]{Shectman2010,Bernstein2014}, all demand detailed turbulence profiling for rigorous site evaluation and optimal AO design. Similarly, the strategic planning for China’s next-generation flagship facilities, such as the proposed Large Optical/Infrared Telescope (LOT; \citealt{Su2017, Feng2020, Cui2018}), relies heavily on high-resolution atmospheric data. This has driven intensive site testing campaigns at prime locations such as Lenghu \citep[selected site for LOT;][]{Deng2021, Yang2024} and the candidate sites for LOT, including Daocheng \citep{Song2020a}, Ali \citep{Liu2020}, and Muztagh Ata \citep{Xu2020}, aiming to ensure these facilities achieve their optimal seeing-limited performance.

The most commonly used instrument for measuring seeing is the Differential Image Motion Monitor \citep[DIMM;][]{Sarazin1990,Tokovinin2002}, which measures the integral of the refractive index structure constant, $C_n^2$, along the line of sight. However, with the advancement of AO technology, high-resolution stratified atmospheric optical turbulence parameters, especially the vertical distribution of the refractive index structure constant $C_n^2(z)$, are of increasing significance. Turbulence profile information is crucial for the design and operation of advanced AO systems, such as Multi-Conjugate Adaptive Optics \citep[MCAO;][]{Beckers1988} and Wide-Field Adaptive Optics \citep[WFAO;][]{LeLouarn2006}.

Over the past decades, a wide range of techniques have been developed to measure either the integrated seeing or the vertical profile of $C_n^2(z)$. To retrieve the vertical distribution of turbulence, profiling instruments utilizing triangulation techniques by observing binary stars were developed, most notably Scintillation Detection and Ranging \citep[SCIDAR;][]{Vernin1973} and Slope Detection and Ranging \citep[SLODAR;][]{Wilson2002}. SCIDAR is a scintillation-based technique, which delivers high vertical resolution and exceptional sensitivity to high-altitude turbulence. In contrast, SLODAR relies on measuring wavefront slopes and can provide high-resolution profiling of the ground layer, especially when deployed in a surface-layer configuration. The Multi-Aperture Scintillation Sensor \citep[MASS;][]{Kornilov2003} complements DIMM by measuring turbulence in discrete high-altitude layers \citep{Kornilov2007, Tokovinin2007}. Other ground-layer techniques include Sonic Detection and Ranging \citep[SODAR;][]{Peters1978}, Shadow-Band Ranger \citep[SHABAR;][]{Moore2006,Sliepen2010}, Generalized SCIDAR \citep[G-SCIDAR;][]{Fuchs1998, Klueckers1998}, Lunar Scintillometer \citep{Hickson2004,Rajagopal2008}, and the Profiler of Differential Solar Limb \citep[PDSL;][]{Song2020b}. Each method has its own strengths and limitations, and they are in many ways complementary.

Despite these advances, most existing methods require specialized instrumentation. This limits their portability and deployment flexibility, especially in extreme environments. To address these issues, \citet{Hickson2019} proposed the Multistar Turbulence Monitor (MTM), a novel method to infer the vertical distribution of turbulence using differential image motion between multiple stars across a wide field of view. By analyzing a sequence of short-exposure images of a star field, differential motion between all pairs of stars is used to compute the structure functions of longitudinal and transverse wavefront tilt for a range of angular separations. The method estimates the turbulence profile, as well as the integrated seeing $\varepsilon$ and outer scale $L_0$, by fitting theoretical models through a Markov Chain Monte Carlo (MCMC) inversion. The method was successfully verified through simulations and preliminary data from the second of the three Antarctic Survey Telescopes (AST3-2) at Dome A in Antarctica.

However, the AST3-2 is a specialized 0.5-meter telescope designed for photometry in Antarctica, making it unsuitable for portable site testing tasks. The MTM method has potential for broader application. The simplicity of its observational requirements suggests that MTM could be implemented with a small-aperture telescope equipped with a fast, wide-field camera. Such components are readily available commercially. Since small telescopes are cost-effective and easier to deploy than specialized monitors, their use as turbulence monitors offers a flexible solution for assessing seeing quality, especially at potential astronomical sites with no existing facilities.

In this work, we present a comprehensive investigation of the performance of the MTM method with different telescopes through theoretical and simulation analyses, and to verify it using actual observational data from the Daocheng Astronomical Site. To achieve this, we first analyze the theoretical basis of MTM and evaluate its requirements for key instrumental parameters, such as aperture and field of view. Subsequently, we simulate the detection capability of the MTM method under typical small-aperture telescope configurations. These analyses provide the foundation for the on-sky validation presented in the following sections.

\section{Theory}

\begin{figure}
    \centering
    \includegraphics[width=0.9\linewidth]{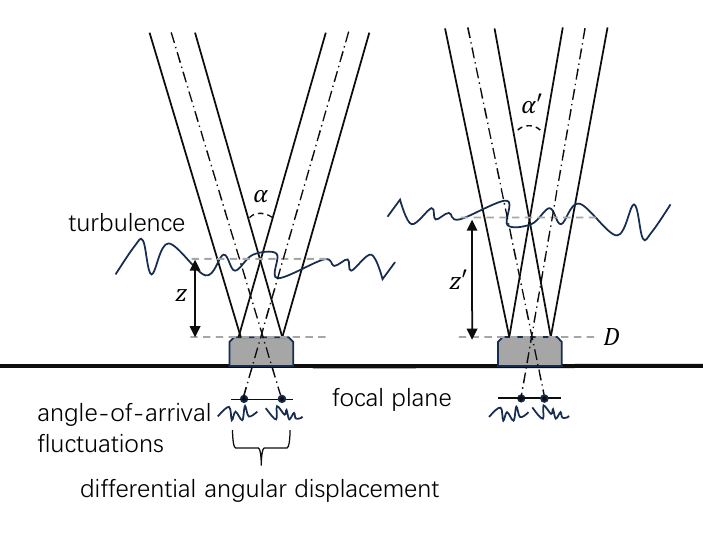}
    \caption{Schematic diagram of the MTM principle. The panels illustrate the optical paths for stellar pairs with different angular separations $\alpha$. Light traverses different paths through the turbulent atmosphere, entering a telescope with aperture diameter $D$. The degree of overlap between the projected telescope pupils varies with height depending on the angular separation $\alpha$. The turbulence-induced wavefront distortions result in differential angle-of-arrival fluctuations, which are measured as the relative displacement of the stellar centroids in the focal plane.}
    \label{fig:MTM}
\end{figure}

\begin{figure*} 
  \centering
  \begin{subfigure}{0.48\textwidth}
    \includegraphics[width=\linewidth]{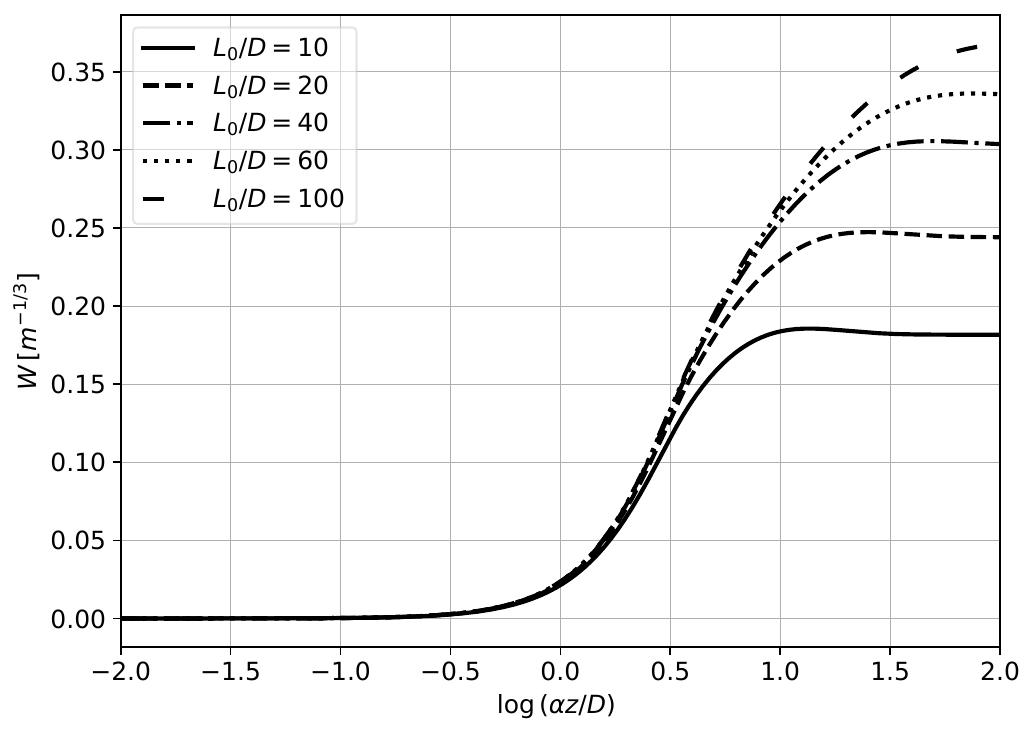}
  \end{subfigure}
  \hfill 
  \begin{subfigure}{0.48\textwidth}
    \includegraphics[width=\linewidth]{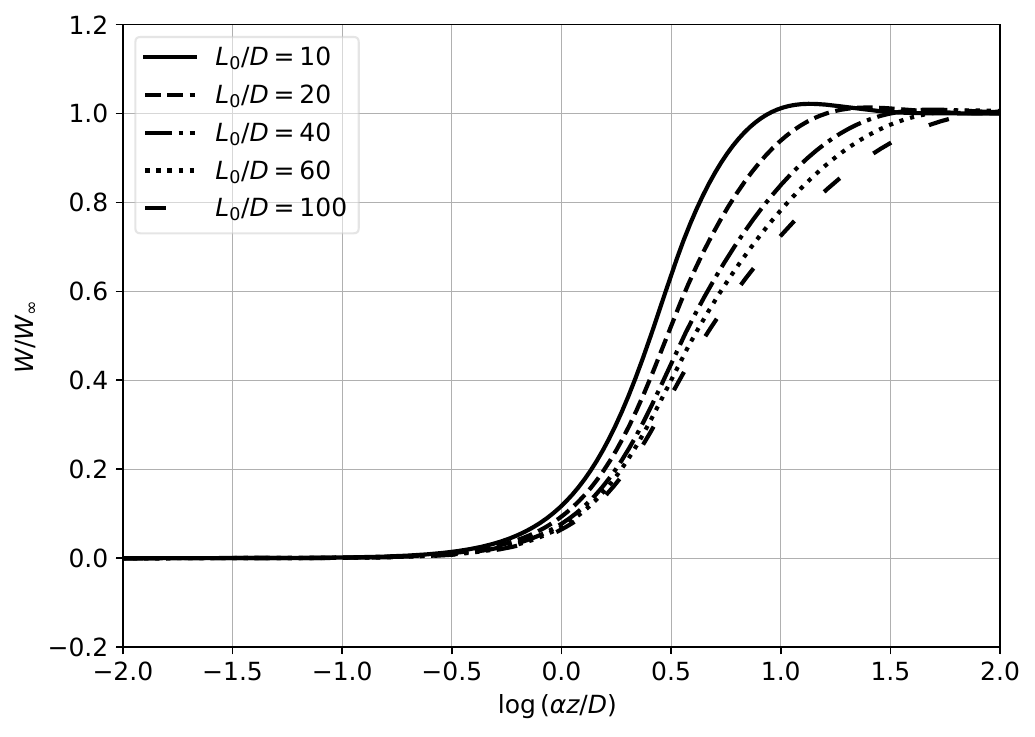}
  \end{subfigure}
  \caption{Response functions $W(\alpha, z)$ calculated for different ratios of the outer scale to the aperture diameter, $L_0/D$. The left panel shows the function value, while the right panel displays the function normalized by its asymptotic value at large separations ($W_{\infty}$). The horizontal axis represents the normalized parameter $\alpha z/D$. The response function indicates the sensitivity of the differential image motion variance to turbulence at a height $z$ for a given angular separation $\alpha$.}
  \label{fig:W and W/W_infty}
\end{figure*}

\subsection{Measurement principle}

The principle of the MTM is illustrated in \cref{fig:MTM}. The diagrams depict the optical propagation paths for stellar pairs with different angular separations $\alpha$. When light from two stars separated by an angular separation $\alpha$ propagates through the atmosphere to the telescope aperture of diameter $D$, the beams traverse distinct optical paths. Atmospheric turbulence induces wavefront distortions that result in angle-of-arrival (AA) fluctuations, causing the centroids of the stellar images to jitter randomly in the focal plane.

The ability of MTM to resolve the vertical turbulence profile stems from the geometric correlation of these optical paths. For a stellar pair with separation $\alpha$, the physical separation between the beams increases with height, causing the degree of overlap between the projected telescope pupils to vary with height. Turbulence layers at different heights are thus probed with different weights depending on this overlap. By measuring the variance of the differential motion for a large number of stellar pairs with varying angular separations, the turbulence strength at different heights can be reconstructed.

This approach shares similarities with the Differential Image Motion Monitor (DIMM; \citealt{Sarazin1990}; \citealt{Tokovinin2002}), yet with a fundamental distinction. A standard DIMM typically utilizes a mask with two sub-apertures, with a prism placed over one sub-aperture to produce twin images of a single star. This configuration is optically equivalent to observing a source with a fixed physical beam separation along the entire line of sight. As a result, DIMM provides only an integrated measure of turbulence and cannot resolve its vertical distribution. In contrast, MTM utilizes the full aperture to measure differential motion between distinct stars. Since the physical beam separation varies linearly with height, MTM effectively probes the atmosphere with height-dependent sensitivity, enabling the retrieval of the $C_n^2(z)$ profile using a standard, portable telescope.

Despite these differences, MTM retains the primary advantage of the differential method. Since global telescope vibrations and tracking errors affect all stars identically, analyzing the relative motion between centroids effectively cancels these common-mode noise sources, ensuring that the measured signal is dominated by atmospheric turbulence. It should be explicitly noted that the theoretical formulation presented in this section, as well as our subsequent empirical data processing pipeline, assumes an instantaneous exposure time ($t \to 0$), thereby neglecting the temporal averaging effects on the measured differential variances.

\subsection{Mathematical formalism}

The theoretical derivation of the differential image motion variance for MTM is detailed in \citet{Hickson2019}. Here, we provide a summary of the mathematical formalism used for the turbulence profile reconstruction.

It is necessary to clearly define the geometric notation used in the following equations. Let $h$ denote the vertical height above the observatory and $z$ denote the propagation distance along the line of sight (range). For observations at a zenith angle $\zeta$, these quantities are related by $z = h \sec \zeta$. To simplify the mathematical representation in this section, we assume zenith viewing ($\zeta = 0^\circ$), such that the distance $z$ is equivalent to the vertical height $h$.

Atmospheric refractive-index fluctuations are modeled assuming isotropic turbulence following the von K\'arm\'an power spectrum:
\begin{equation}
    \Phi_n(\kappa) = 0.033 C_n^2(z) (\kappa^2 + \kappa_0^2)^{-11/6},
	\label{eq:spectrum}
\end{equation}
where $\kappa$ is the magnitude of the spatial frequency vector, $\kappa_0 = 2\pi/L_0$, and $L_0$ is the outer scale of turbulence.

Following the integration over the telescope aperture $D$, the total structure function is related to the turbulence profile $C_n^2(z)$ by:
\begin{equation}
	\D(\alpha) = \int_{0}^{\infty} C_n^2(z) W(\alpha, z) dz.
	\label{eq:integral_form}
\end{equation}
where $\alpha$ is the angular separation. Here, $W(\alpha, z)$ is the response function (or weighting kernel), defined as:
\begin{equation}
    \begin{split}
    W(\alpha, z) &=  \, 41.68 \, D^{-1/3} \\
    &\quad \times \int_0^{\infty} \kappa (\kappa^2 + \kappa_0^2)^{-11/6} J_1^2\left(\frac{\kappa D}{2}\right) [1 - J_0(\kappa z \alpha)] \, d\kappa.
    \end{split}
    \label{eq:weighting_function}
\end{equation}
where $J_0$ and $J_1$ are Bessel functions.

Equation \ref{eq:integral_form} establishes the linear relationship between the observable differential variance $\D(\alpha)$ and the unknown turbulence profile $C_n^2(z)$. Since the response function $W(\alpha, z)$ depends on distance $z$, measuring $\D(\alpha)$ at multiple angular separations $\alpha$ allows for the inversion of the integral equation to reconstruct $C_n^2(z)$.

\section{METHOD}
\subsection{Response function}

The inversion of the turbulence profile relies on the distinct behavior of the response function $W(\alpha, z)$ at different heights and angular separations. Since the telescope aperture $D$ is fixed for a given instrument, $W(\alpha, z)$ can be pre-calculated for various combinations of $\alpha$, $z$, and the outer scale parameter $L_0$.

Figure \ref{fig:W and W/W_infty} displays examples of the response function $W(\alpha, z)$ for several values of the ratio $L_0/D$. The response magnitude depends primarily on the dimensionless parameter $\alpha z / D$, which represents the ratio of the physical separation of the optical paths at height $z$ to the telescope aperture diameter $D$.

As shown in \cref{fig:W and W/W_infty}, $W(\alpha, z)$ increases monotonically with $\alpha z / D$ before saturating to a constant value. Crucially, the ability to reconstruct the vertical profile depends on sampling the non-saturated regime of these curves, where the sensitivity to height variations is maximized, rather than the saturated region where the weighting is uniform. For high-altitude turbulence (large $z$), the response function saturates rapidly even at small separations; therefore, data at small $\alpha$ are essential to probe the steep rise of the response function. Conversely, for the ground layer (small $z$), the response remains negligible until the angular separation becomes sufficiently large. Thus, capturing the contribution of near-ground turbulence requires measurements at large angular separations. This height-dependent geometric correlation is precisely what enables the MTM to retrieve the vertical turbulence profile $C_n^2(z)$.

\begin{table*}
	\centering
	\caption{Key instrumental parameters of the two contrasting MTM configurations used in this study.}
	\label{tab:instrument_params}
	\begin{tabular}{lcccccccc} 
		\hline
		Configuration & Telescope & Aperture & Focal Ratio & Camera & Pixel Size & Pixel Scale & Resolution & FOV\\
		& & (mm) & & & ($\mu$m) & ($''$/pix) & (pixels) & ($'$)\\
		\hline
		Wide-field & RASA 11 & 279 & f/2.2 & QHY268M & 3.76 & 1.25 & $6280 \times 4210$ & $130.9 \times 87.8$\\
		High-resolution & C925 HD & 235 & f/10 & ASI1600MM & 3.80 & 0.33 & $4656 \times 3520$ & $25.9 \times 19.6$\\
		\hline
	\end{tabular}
\end{table*}

\subsection{Inversion technique}
In practice, the estimation of the turbulence profile involves two stages: calculating the observed structure function from image sequences and subsequently inverting the integral equation.

First, stellar centroids are extracted from the raw short-exposure images. For each star $i$ in a given frame, the centroid position $\mathbf{x}_i$ is determined using a centroiding algorithm. The mean centroid position over the sequence is denoted as $\bar{\mathbf{x}}_i$, and the angular separation between any two stars is $r_{ij} = ||\bar{\mathbf{x}}_i - \bar{\mathbf{x}}_j||$. The variance of the differential image motion for each pair constitutes the observed structure function $\mathcal{D}_{\text{obs}}(\alpha)$.

Since direct inversion of the Fredholm integral in Eq. (\ref{eq:integral_form}) to retrieve the turbulence profile $C_n^2(z)$ is an ill-posed problem sensitive to noise, we employ a forward modeling approach. The vertical distribution of optical turbulence is parameterized by a finite set of model parameters $\mathbf{m} = \{m_1, m_2, \dots, m_n\}$, representing the $C_n^2$ values at discrete height nodes and the outer scale $L_0$. In the forward modeling step, the theoretical structure function is calculated by approximating the continuous integral in Eq. (\ref{eq:integral_form}) as a discrete summation, where the $C_n^2$ values are explicitly multiplied by the corresponding integration step size (layer widths) $\Delta z$. Therefore, the inverted parameters represent the localized profile $C_n^2(z)$ (in units of $\rm{m}^{-2/3}$).

We utilize a Markov Chain Monte Carlo (MCMC) sampling method to find the parameter set that minimizes the discrepancy between the theoretical model and observations. In each iteration, the model parameters are perturbed, and the goodness-of-fit is evaluated using a chi-square statistic:
\begin{equation}
\chi^2 = \sum_{k} \frac{[\mathcal{D}_{\text{obs}}(\alpha_k) - \mathcal{D}_{\text{mod}}(\alpha_k; \mathbf{m})]^2}{\sigma_k^2},
\label{eq:chi_square}
\end{equation}
where the summation runs over all distinct stellar pairs (indexed by $k$). Here, $\sigma_k$ represents the statistical uncertainty of the measured differential image motion variance $\mathcal{D}_{obs}(\alpha_k)$ for the $k$-th pair. The MCMC process samples the posterior probability distribution, providing not only the best-fit $C_n^2(z)$ profile but also robust uncertainty estimates for all parameters. 

Once the optimal parameters are determined, the resulting $C_n^2(z)$ profile is integrated to derive the cumulative turbulence strength. Specifically, the Fried parameter $r_0$ and the full width at half maximum (FWHM) of the seeing disk $\varepsilon$ are computed as \citep{Roddier1981}:
\begin{equation}
    r_0 = \left[ 0.423 k^2 \int_0^{\infty} C_n^2(z) \, dz \right]^{-3/5},
    \label{eq:r0}
\end{equation}
\begin{equation}
    \varepsilon = 0.98 \frac{\lambda}{r_0},
    \label{eq:seeing}
\end{equation}
where $k = 2\pi/\lambda$ is the optical wavenumber and $\lambda$ is the wavelength.

\section{INSTRUMENTAL CONFIGURATIONS AND REQUIREMENTS}

\subsection{Key instrumental requirements}
\label{sec:requirements}
The MTM technique relies on measuring the differential image motion of multiple stars within sequences of short-exposure frames. Therefore, the instrument must simultaneously provide a sufficiently wide field of view (FOV) to include multiple stars, a large aperture to increase the number of detectable stars and enhance the signal-to-noise ratio (SNR), and a fine pixel scale to ensure accurate centroid determination. The key instrumental parameters governing the measurement performance are the telescope aperture $D$, focal length $f$, pixel size, and sensor format ($N_x \times N_y$).

The requirement for the FOV is determined by the dynamic range of the response function. As analyzed in Section 3.1, the response magnitude varies significantly with the dimensionless parameter $\log(\alpha z / D)$. For high-altitude turbulence (e.g., $z = 10000$ m), the response saturates rapidly; the saturation point, corresponding to $\log(\alpha z / D) \approx 1.5$, translates to angular separations of no more than $\sim 130''$ for a $D=0.2$ m telescope. Conversely, to effectively probe the ground layer (e.g., $z \approx 30$ m), the measurement must cover the region where the response function enters its dynamic variation range (sensitive zone), which typically begins around $\log(\alpha z / D) \approx 0$. This implies that angular separations must exceed $20'$. Consequently, an optimal MTM instrument must possess a wide FOV to sample these large angular separations.

Regarding the telescope aperture, a larger diameter improves photon collection, allowing for the detection of fainter stars and thus increasing the number of available stellar pairs. Based on prior on-site observations at Daocheng with an exposure time of $t = 50$ ms, the limiting magnitude $m_{\text{lim}}$ in the V-band is estimated empirically as:
\begin{equation}
m_{\text{lim}} \approx C_{\text{site}} + 2.5 \log_{10}(D),
\label{eq:limiting_mag}
\end{equation}
where $D$ is the aperture in meters and $C_{\text{site}} \approx 16.61$ is an empirical zero-point calibration constant derived from our actual measurement data. In this context, the metric used to define the ``limiting magnitude" is the magnitude of the faintest star that achieves a SNR of 5. For a 0.2 m telescope, this yields a limiting magnitude of approximately 14.80 mag. Furthermore, a larger aperture improves the SNR for a given star, which directly enhances centroiding precision ($\sigma_{\text{cent}} \propto \text{FWHM}/\text{SNR}$). However, while increasing the aperture improves performance, the primary objective of MTM is to measure atmospheric seeing with compact instrumentation. Therefore, keeping the telescope aperture below 0.3 m strikes an optimal balance between performance and portability.

Centroiding accuracy is critical for measuring differential image motion and is fundamentally limited by SNR and pixel sampling. For typical small-aperture telescopes, atmospheric turbulence breaks the short-exposure star image into speckle patterns whose structure is governed by the $D/r_0$ ratio. To avoid undersampling and minimize aliasing, these short-exposure speckles must be adequately sampled. As a general practical rule, the Full Width at Half Maximum (FWHM) of the speckles should span at least 2 to 3 pixels. Since the FWHM of an individual speckle is dictated by the diffraction limit ($1.22\lambda/D$), which is approximately $0.4^{\prime\prime}$ to $0.5^{\prime\prime}$ for typical small-aperture telescopes at visible wavelengths, achieving optimal sampling requires a pixel scale finer than approximately $0.25^{\prime\prime}$ per pixel. However, achieving such fine pixel scales inevitably requires a longer focal length, which reduces the FOV for a given detector size. The conflict arises between these high-resolution requirements and the need for a wide FOV. A short-focal-length system provides the wide FOV essential for ground-layer profiling but results in coarser sampling, which may degrade sensitivity to high-altitude turbulence where differential motions are minute. Conversely, a long-focal-length system ensures excellent sampling accuracy but often lacks the FOV to fully resolve the ground layer. Therefore, practical MTM implementations must carefully balance these competing factors.

\subsection{Selected instrumental configurations}
\label{sec:configurations}
To evaluate the MTM performance under realistic conditions and investigate the trade-offs discussed above, two distinct commercially available optical systems were selected for simulation and on-sky testing. These configurations represent different optimization strategies within the parameter space of portable monitors. The key specifications of these two systems are summarized in Table \ref{tab:instrument_params}.

\textbf{(i) Wide-field configuration: Celestron RASA 11 + QHY268M.} \\
This setup utilizes the Celestron RASA 11 telescope ($f/2.2$). Paired with the large-format QHY268M camera, it delivers an exceptionally wide field of view ($\sim 131' \times 88'$). The APS-C sensor format (diagonal $\sim 28.3$ mm) of the QHY268M camera falls well within the highly corrected, flat image circle (diameter $43.3$ mm) of the RASA 11 optical design. This wide coverage allows for the measurement of angular separations large enough to probe turbulence layers as low as $\sim 10$ m above the ground (based on the $\alpha \approx D/z$ relation). However, the short focal length results in a coarser pixel scale ($\sim 1.25''$/pixel), leading to undersampling and lower centroiding precision. Consequently, while this configuration excels at ground-layer profiling, it has reduced sensitivity to the minute differential motions induced by high-altitude turbulence.

\textbf{(ii) High-resolution configuration: Celestron C925 HD + ZWO ASI1600MM.} \\
This setup employs the Celestron C925 HD ($f/10$) coupled with the ZWO ASI1600MM camera. This combination yields a fine pixel scale ($\sim 0.33''$/pixel), ideal for detecting the subtle differential motions caused by high-altitude turbulence. However, the narrower FOV ($\sim 26' \times 20'$) limits the number of measurable stellar pairs and restricts the maximum accessible separation angle $\alpha$. Therefore, while inherently more sensitive to high-altitude turbulence, this configuration may struggle to fully resolve the ground layer structure.

Overall, the choice of telescope and detector for practical MTM applications involves a complex trade-off between light-collecting power, field of view, spatial sampling, and portability. While an ideal instrument would simultaneously maximize all these parameters, practical constraints often require prioritizing specific capabilities. These two extreme configurations are adopted for the detailed simulations and on-sky observations presented in the following sections. This approach allows us to validate the robustness of the MTM inversion algorithm across different sampling regimes and verify that the derived turbulence parameters are instrument-independent.

\begin{figure*} 
  \centering

  \begin{subfigure}{0.49\textwidth} 
    \includegraphics[width=\linewidth]{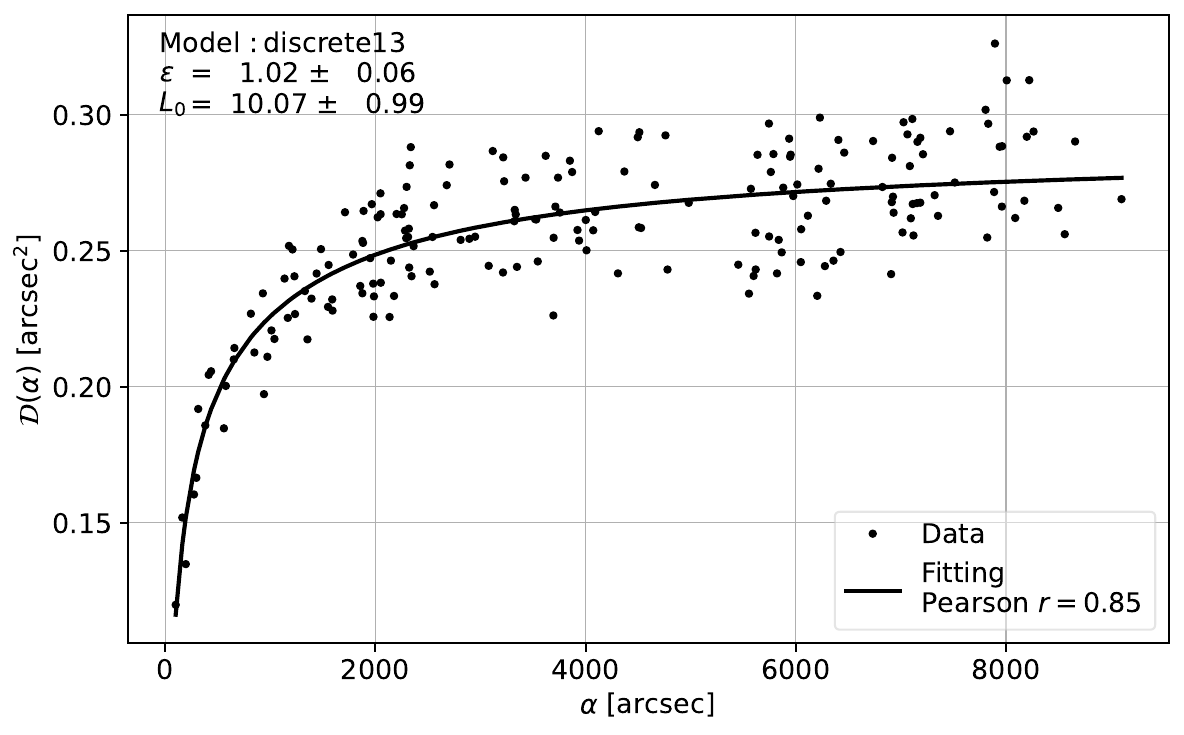} 
  \end{subfigure}
  \begin{subfigure}{0.49\textwidth}
    \includegraphics[width=\linewidth]{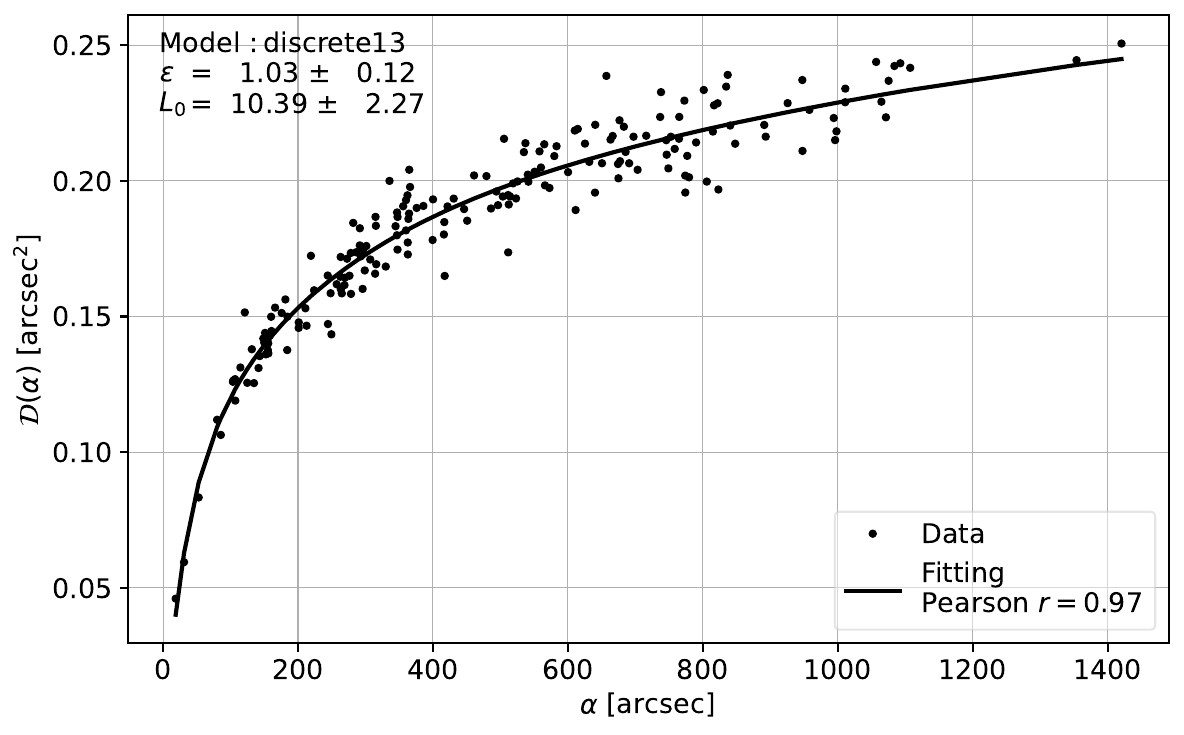}
  \end{subfigure}
  

  \begin{subfigure}{0.36\textwidth}
    \includegraphics[width=\linewidth]{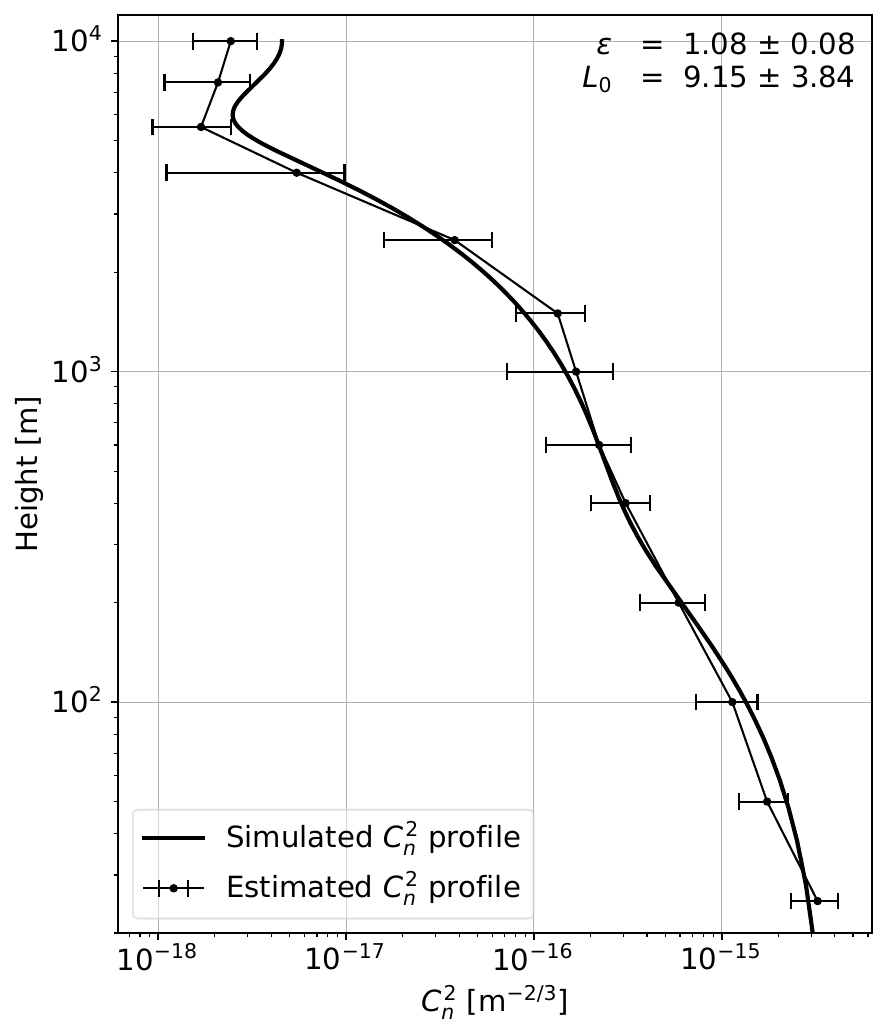}
  \end{subfigure}
  \begin{subfigure}{0.36\textwidth}
    \includegraphics[width=\linewidth]{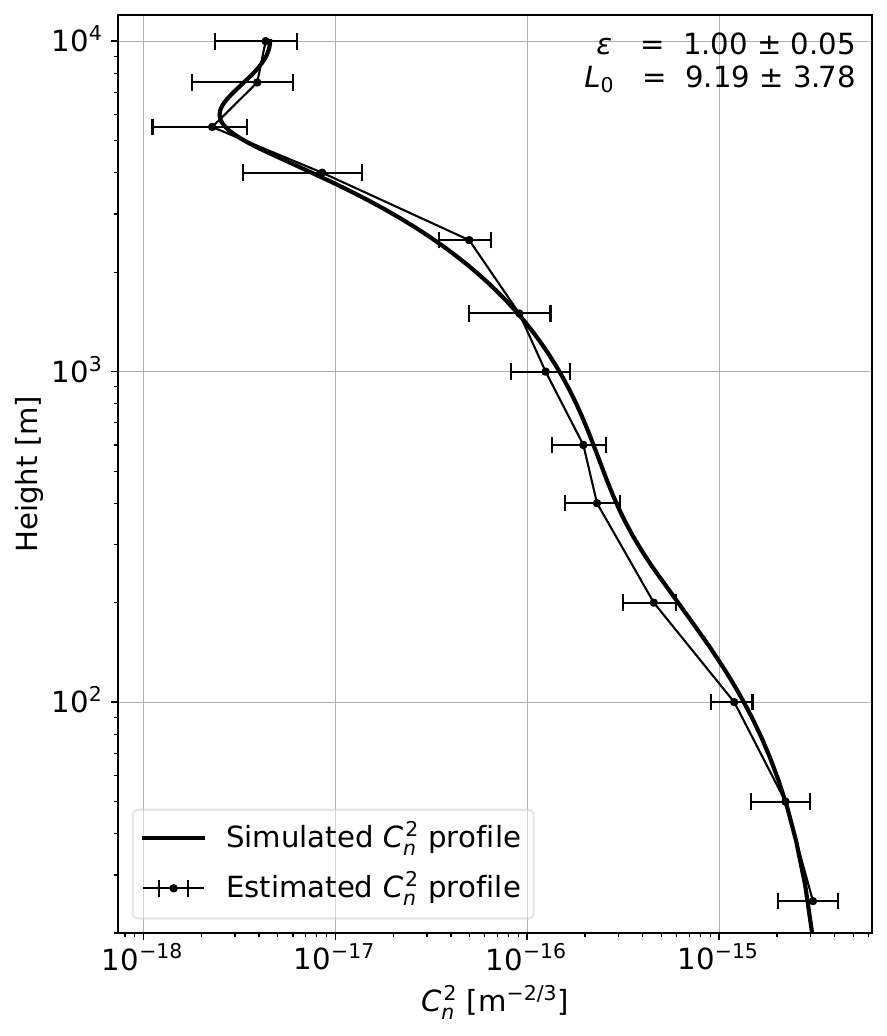}
  \end{subfigure}
  \caption{
Simulation-based inversion results for the RASA 11 (left column) and C925 HD (right column) configurations. Top row: Fitting of the differential image motion variance structure functions for a single realization. Bottom row: Statistical results of the retrieved turbulence profiles from 30 independent inversions. The solid black lines represent the input HV model ($\varepsilon = 1.0''$, $L_0 = 10$ m). The points with error bars indicate the mean estimated $C_n^2$ values and their standard deviations at the 13 discrete height nodes. Both systems recover the integrated parameters well, though the high-resolution configuration shows better constraints on high-altitude layers.
}
  \label{fig:Simulation}
\end{figure*}

\section{Simulations}

\subsection{Simulated data generation}

To assess the viability and performance of the MTM inversion technique, synthetic observational data were generated based on a representative turbulence profile. This profile was constructed using the standard Hufnagel--Valley (HV) model \citep{Hufnagel1964, Valley1980}, parameterized with a total seeing of $1.0''$ and an outer scale of $L_0 = 5$ m.

The simulations were performed using the two instrumental configurations defined in Section \ref{sec:configurations}: the wide-field RASA 11 system and the high-resolution C925 HD system. To ensure the simulation reflects realistic observing conditions, rather than using a random stellar distribution, we adopted the actual celestial coordinates of stars within the open cluster M34 (NGC 1039). M34 was selected as a representative target because it exemplifies the ideal characteristics for MTM observations: it is a loose open cluster providing a sufficient density of bright stars to sample a wide range of angular separations. Furthermore, the cluster culminates near the zenith at the Daocheng site during the winter.

From the cluster catalog, the 20 brightest stars were selected to represent high-SNR targets. This sample size yields $N(N-1)/2 = 190$ independent pairs, ensuring a dense sampling of angular separations for the structure function analysis. For each instrumental configuration, 100 statistically independent frames were generated, explicitly assuming an instantaneous exposure time ($t \rightarrow 0$). The theoretical variances of the differential image motion for all stellar pairs were computed from the input HV profile using Eq. (\ref{eq:integral_form}). To generate the synthetic observations, an iterative coordinate-adjustment procedure was applied: the stellar coordinates in each frame were perturbed such that the variance of their differential angular displacements matched the theoretical predictions. Finally, to emulate realistic measurement noise, random centroiding errors were added to the adjusted coordinates in each frame. This injected coordinate error acts as a mathematical equivalent to the total centroiding uncertainty driven by stellar photon shot noise and detector readout noise. These errors were constrained by the pixel sampling of each instrument configuration and scaled according to the brightness of each specific star, reflecting the impact of the signal-to-noise ratio on centroid positioning precision. These noise-perturbed coordinate datasets constitute the final simulated input for the MTM inversion tests. 

\subsection{Analysis of simulated data}

The MCMC inversion pipeline estimates the turbulence profile that best reproduces the simulated differential image-motion statistics, simultaneously yielding the total seeing $\varepsilon$ and the outer scale $L_0$. The framework adopts a discrete-layer representation of the turbulence with fourteen free parameters: the $C_n^2$ values at thirteen discrete height nodes spanning 20 m to 10000 m and the outer scale $L_0$.

The choice of thirteen height nodes represents an optimal compromise between vertical resolution and computational efficiency. While a larger number of nodes would theoretically provide a finer profile, it significantly expands the parameter space. This expansion dilutes the constraints provided by the data, requiring a substantial increase in MCMC iterations to achieve convergence and stable posterior distributions. Conversely, fewer nodes would fail to adequately capture complex atmospheric stratification. With the adopted configuration of 20 stars and sequences of 100 frames, operating on a standard personal laptop equipped with an AMD Ryzen 9 processor (3.30 GHz) and 32 GB of RAM, a typical inversion run of 2000 MCMC iterations requires a computation time of approximately 20 seconds.

Figure \ref{fig:Simulation} presents the inversion results for the two instrumental configurations. The top row displays the fitting of the differential image motion variance as a function of stellar separation $\alpha$. Both the wide-field RASA 11 (left) and high-resolution C925 HD (right) configurations demonstrate excellent agreement between the simulated data (points) and the theoretical structure functions (solid lines) derived from the inverted profiles.

The bottom row illustrates the statistical results of the reconstructed turbulence profiles from 30 independent inversions. The simulated turbulence profile (solid black curve) is constructed using the HV model with $\varepsilon = 1.0''$ and $L_0 = 10$ m. Both configurations robustly recover the integrated seeing and outer scale within the estimated uncertainties. However, the profile reconstruction reveals the impact of instrumental sampling:

\begin{itemize}
    \item \textbf{Wide-field (RASA 11 + QHY268M):} While the wide FOV is theoretically advantageous for the ground layer, the coarser pixel scale ($1.25''$/pixel) introduces undersampling effects. As observed in the left panel, this leads to larger uncertainties (longer error bars) and a slight deviation from the true profile at higher heights compared to the C925 setup. Furthermore, the expected advantage in ground-layer resolution is partially offset by the reduced centroiding precision, resulting in a performance comparable to the C925 system at lower altitudes.
    \item \textbf{High-resolution (C925 HD + ASI1600MM):} Benefiting from its fine pixel scale ($0.33''$/pixel), this system exhibits superior sensitivity to high-altitude turbulence. As shown in the right panel, the reconstructed profile at high heights ($z > 1$ km) closely tracks the theoretical curve with small error bars, demonstrating that fine sampling is critical for resolving the minute differential motions caused by distant layers.
\end{itemize}

These results confirm that while a wide FOV is necessary for ground-layer coverage, sufficient spatial sampling is equally vital for accurate profiling across the entire atmosphere.

\begin{figure}
    \centering
    \includegraphics[width=\columnwidth]{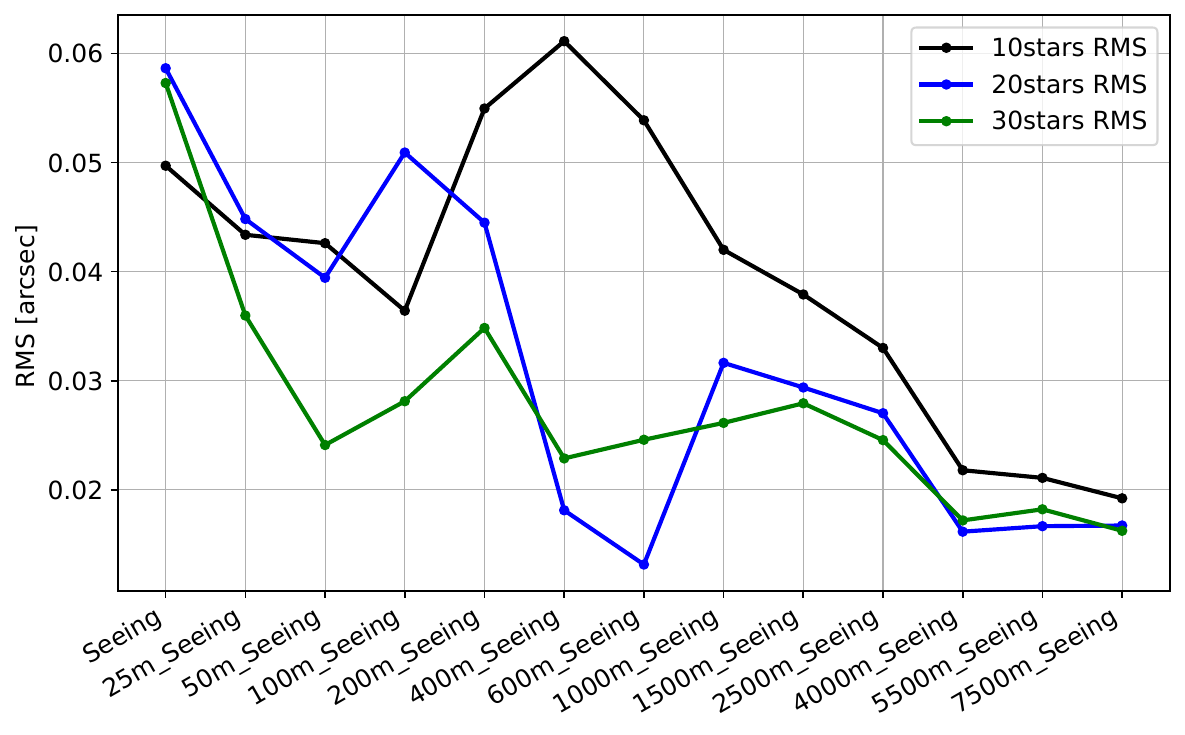} 
    \caption{Simulated RMS error of the retrieved seeing at different discrete heights as a function of the number of stars used in the inversion.}
    \label{fig:star_num}
\end{figure}

To determine the minimum number of stars required for robust MTM inversion, we conducted simulations evaluating the retrieval accuracy as a function of the stellar sample size. Figure \ref{fig:star_num} illustrates the RMS error of the retrieved seeing at discrete heights for configurations using 10, 20, and 30 stars. The results demonstrate that sacrificing the number of usable stellar pairs leads to large reconstruction errors across most height nodes due to insufficient sampling of the spatial structure function. Extracting approximately 20 stars (yielding 190 independent baselines) reduces the RMS error to a stable level, providing sufficient constraints for the MCMC algorithm. Therefore, maintaining a minimum of $\sim20$ stars is a prerequisite for reliable turbulence profiling with this method.

To further quantify the linearity and robustness of the inversion algorithm, extensive simulations were performed across a broad parameter space, covering seeing values from $0.6''$ to $1.8''$ and outer scales from 2 m to 30 m. The retrieval performance is summarized in Table \ref{tab:sim_linearity}. Notably, both instrumental configurations demonstrate consistent accuracy in recovering the integrated atmospheric parameters. The inferred seeing values exhibit only minor deviations across the full dynamic range, effectively tracking the input from excellent ($0.6''$) to poor ($1.8''$) conditions. Similarly, the retrieved outer scale $L_0$ is broadly consistent with the preset parameters, distinguishing clearly between compact ($L_0 \approx 2$ m) and developed ($L_0 \approx 30$ m) turbulence regimes. These results underscore the intrinsic algorithmic stability of the MTM inversion technique across varying turbulence strengths. However, these simulations are highly idealized, assuming an instantaneous exposure without wind speed or optical propagation effects. Under actual diverse atmospheric conditions, finite exposure times combined with high-altitude wind speeds will lead to a certain degree of systematic bias.

\begin{table}
	\centering
	\caption{Summary of the retrieval accuracy for integrated turbulence parameters under varying atmospheric conditions. The table compares the input model values with the median results obtained from 30 independent simulation runs for both the wide-field and high-resolution configurations.}
	\label{tab:sim_linearity}

	\begin{tabular}{lcccc} 
		\hline
		\multicolumn{1}{c}{\textbf{Input}} & \multicolumn{2}{c}{\textbf{Wide-field}} & \multicolumn{2}{c}{\textbf{High-resolution}} \\ 
		$\varepsilon$\hspace{12pt}$L_0$&$\varepsilon$&$L_0$&$\varepsilon$&$L_0$\\
        ($''$) \hspace{2pt} (m) & ($''$) & (m) & ($''$) & (m) \\
		\hline
        0.6 \hspace{5pt} 2 & 0.64 $\pm$ 0.01& 1.83 $\pm$ 0.79& 0.61 $\pm$ 0.02 & 2.03 $\pm$ 0.22\\
        0.6 \hspace{3pt} 10 & 0.60 $\pm$ 0.02& 13.16 $\pm$ 3.34& 0.60 $\pm$ 0.02& 9.77 $\pm$ 3.69\\
        0.6 \hspace{3pt} 30 & 0.63 $\pm$ 0.01& 27.23 $\pm$ 5.95& 0.59 $\pm$ 0.01 & 24.87 $\pm$ 8.42\\
        
        1.0 \hspace{5pt} 2 &1.02 $\pm$ 0.03& 1.80 $\pm$ 0.71& 1.03 $\pm$ 0.05 & 1.71 $\pm$ 0.52\\
        1.0 \hspace{3pt} 10 & 1.08 $\pm$ 0.08 & 9.15 $\pm$ 3.84 & 1.00 $\pm$ 0.05& 9.19 $\pm$ 3.78\\
        1.0 \hspace{3pt} 30 & 1.03 $\pm$ 0.03& 29.29 $\pm$ 9.38& 1.03 $\pm$ 0.05& 28.67 $\pm$ 11.01\\
       
        1.4 \hspace{3pt} 10 & 1.34 $\pm$ 0.06& 15.65 $\pm$ 7.16& 1.39 $\pm$ 0.05& 10.02 $\pm$ 2.45\\
        1.4 \hspace{3pt} 30 & 1.46 $\pm$ 0.03& 29.83 $\pm$ 8.82& 1.51 $\pm$ 0.06& 28.61 $\pm$ 15.16\\
        
        1.8 \hspace{3pt} 10 & 1.70 $\pm$ 0.04& 16.92 $\pm$ 5.56& 1.90 $\pm$ 0.11& 9.81 $\pm$ 3.31\\
        1.8 \hspace{3pt} 30 & 1.77 $\pm$ 0.04& 32.48 $\pm$ 10.01& 1.98 $\pm$ 0.12& 30.86 $\pm$ 13.52\\
		\hline  
	\end{tabular}

\end{table}

\section{RESULTS AND DISCUSSION}

\subsection{Observational data and strategy}

The observational data presented in this study were obtained at the Mt. Wumingshan Station 1 (Longitude $100^\circ 04' 12''$ E, Latitude $29^\circ 09' 01''$ N) of the Daocheng Astronomical Site in Sichuan, China. This site has been characterized as a prime candidate for future large-aperture facilities \citep{Song2020a}. The campaign was conducted on the nights of 5, 7, and 8 January 2026.

\begin{figure}
    \centering
    \includegraphics[width= 1\linewidth]{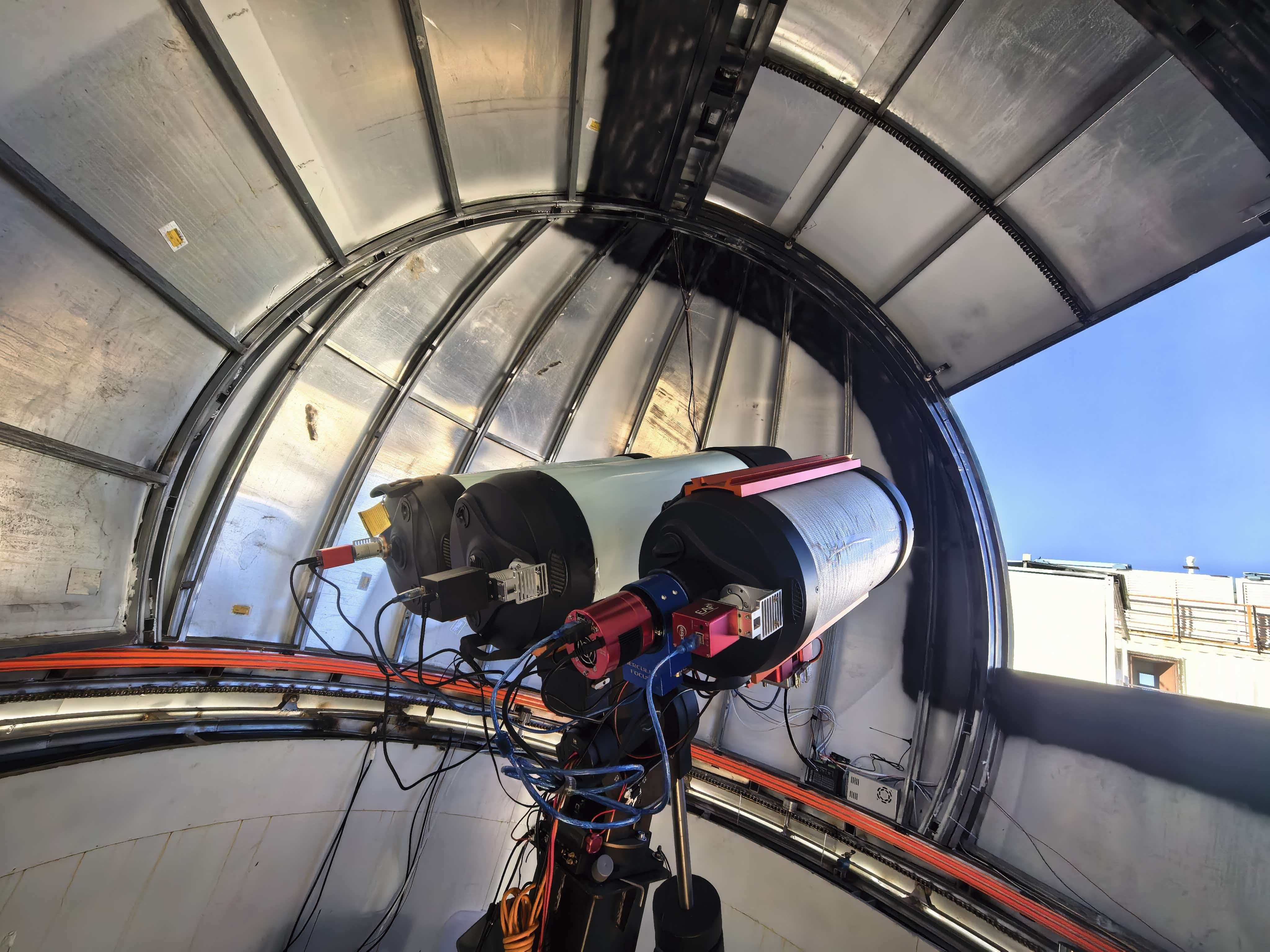}
    \caption{The MTM instruments deployed at the Daocheng Astronomical Site. For this study, only two configurations were utilized: the central telescope is RASA 11 (with the QHY268C camera), and the telescope on the right is C925 HD (with the ZWO ASI1600MM camera). The telescope on the far left was not used. Both telescopes are installed on the same equatorial mount to ensure synchronous tracking.}
    \label{fig:site_instruments}
\end{figure}

We deployed the two MTM configurations detailed in Section \ref{sec:configurations}: the wide-field RASA 11 system and the high-resolution C925 HD system. Figure \ref{fig:site_instruments} shows the on-site deployment of the MTM system and specific parameters are listed in Table \ref{tab:instrument_params}. The reference DIMM instrument is not pictured in Figure \ref{fig:site_instruments}. It is physically located on an adjacent observation tower approximately 20 meters away and 10 meters higher than the MTM dome. On the night of 5 January, both MTM systems operated simultaneously alongside a standard DIMM to provide a rigorous dataset for cross-validation. We obtained one measurement every 10-20 minutes because we need to coordinate the two systems which have different readout time/speed and focusing time. Subsequent observations on 7 and 8 January were carried out using the C925 HD system simultaneously with the DIMM. These runs achieved a higher sampling cadence of approximately 2 minutes, enabling detailed characterization of rapid turbulence evolution.

The observation strategy involved tracking bright star clusters starting from near-zenith positions. To minimize airmass effects, targets were tracked until their elevation dropped below $60^\circ$, at which point the telescope was slewed to a new bright star cluster located near the zenith. It is important to note that the MTM inversion algorithm explicitly incorporates a slant-range correction ($z = h \sec \zeta$); thus, the derived vertical profiles represent the equivalent zenith turbulence distribution regardless of the observing elevation. All data were acquired with an exposure time of 50 ms. We acknowledge that an exposure time of $\le 10$~ms is theoretically optimal to freeze atmospheric turbulence. However, the 50~ms exposure was a practical compromise necessary to robustly validate the method's feasibility on portable equipment. During this campaign, the RASA 11 system was actually equipped with a color CMOS camera (QHY268C), which suffers from a significantly lower quantum efficiency compared to monochrome sensors (QHY268M). Reducing the exposure time would not only decrease the number of detectable stars below this critical threshold but also severely degrade the SNR, drastically increasing centroiding errors. To ensure that both instrument setups could synchronously capture sufficient bright stars, 50~ms was a robust choice. Each turbulence profile and seeing measurement was derived from a sequence of 100 continuous frames.

To process the observational data, we utilized the \textsc{SExtractor} software \citep{Bertin1996} for star detection and centroid extraction from the short-exposure image sequences. The centroid positions were determined using its windowed centroiding algorithm, which yields the highly accurate \texttt{XWIN\_IMAGE} and \texttt{YWIN\_IMAGE} coordinates. For source extraction, the detection threshold was set to $3\sigma$ above the local background noise, while the analysis threshold was set to $2.5\sigma$. These relative thresholding parameters remained strictly identical for both the wide-field RASA 11 and high-resolution C925 HD configurations.

\subsection{Seeing validation and comparison with DIMM}
\label{section:seeing DIMM}

\begin{figure*} 
  \centering

  \begin{subfigure}{1\textwidth} 
    \includegraphics[width=\linewidth]{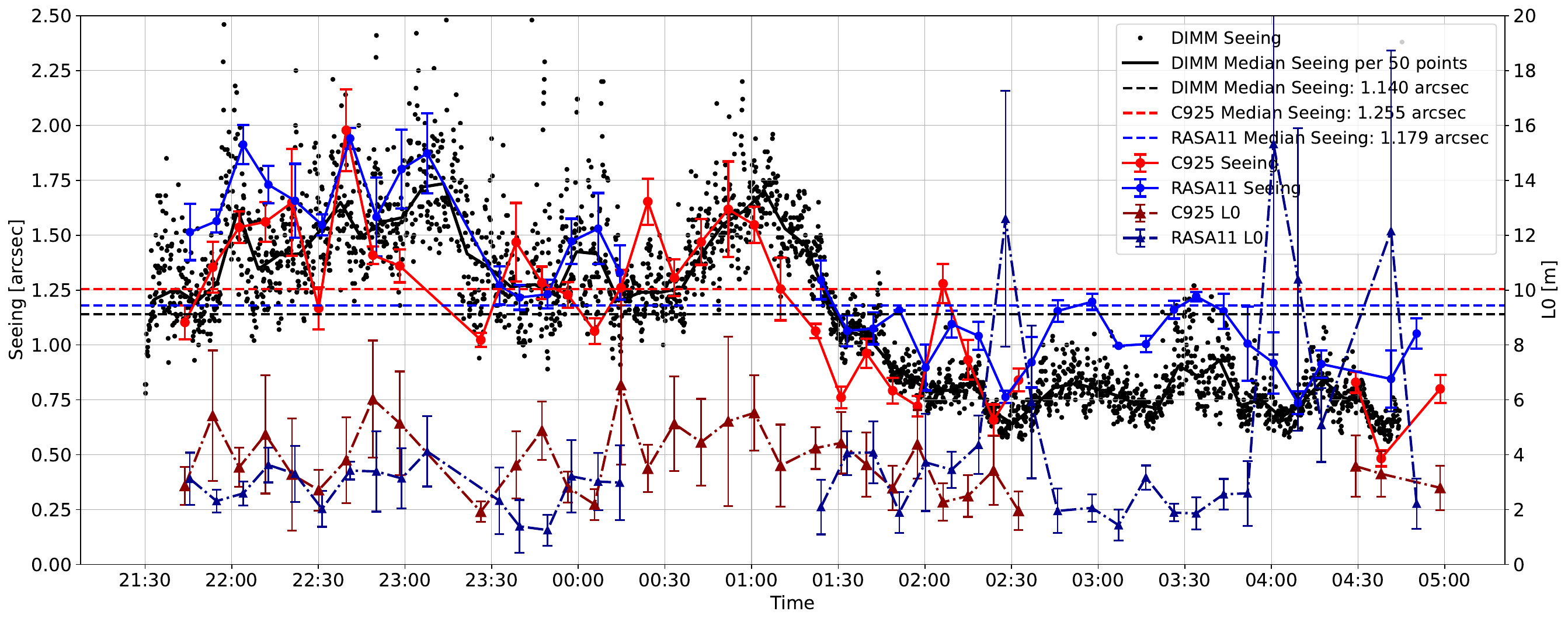} 
  \end{subfigure}

  \begin{subfigure}{1\textwidth}
    \includegraphics[width=\linewidth]{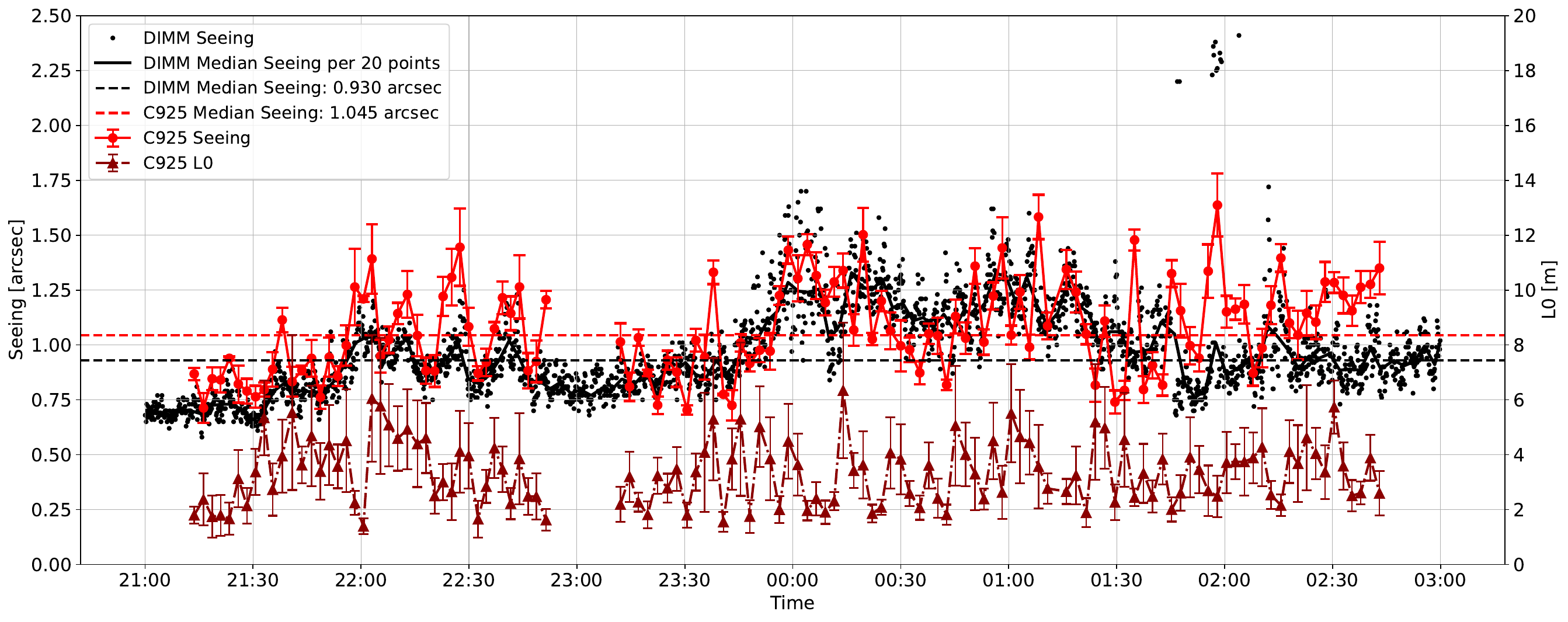}
  \end{subfigure}

  \begin{subfigure}{1\textwidth}
    \includegraphics[width=\linewidth]{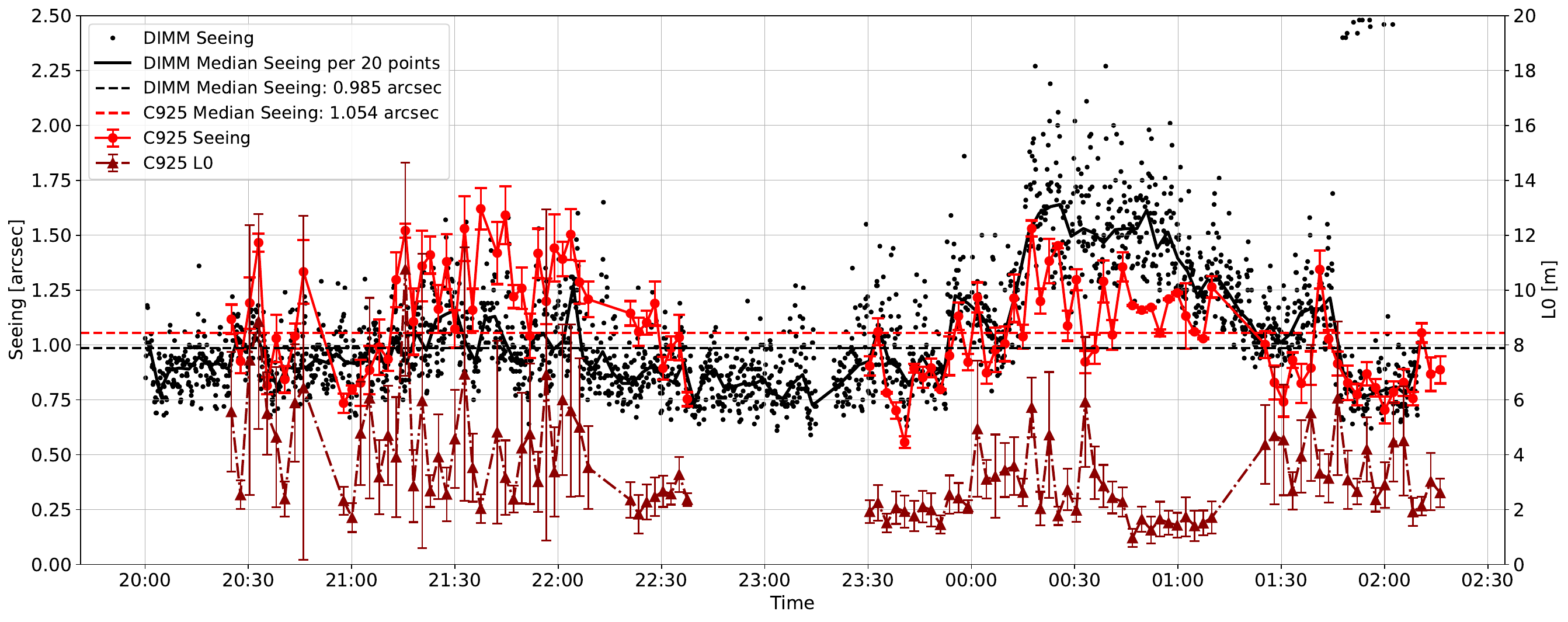}
  \end{subfigure}
  
  \caption{Temporal evolution of atmospheric seeing measured at the Daocheng site. The upper panel shows the simultaneous observations on 5 January, comparing the wide-field RASA 11 (blue line) and high-resolution C925 HD (red line) results with the site-monitoring DIMM (black dots). The corresponding $L_0$ measurements are plotted on the secondary y-axis (dark blue and dark red dash-dot lines). Error bars on the MTM data points represent the $1\sigma$ uncertainties derived from the MCMC posterior distributions. The middle and bottom panels display the continuous monitoring results from the C925 HD system on 7 and 8 January, respectively. Gaps in the time series indicate periods of invalid data due to cloud cover or extreme wind conditions.}
  \label{fig:seeing_evolution}
\end{figure*}

The temporal evolution of the atmospheric seeing measured throughout the three nights is presented in Figure \ref{fig:seeing_evolution}. The upper panel displays the simultaneous results from the RASA 11 and C925 HD systems compared with the DIMM on 5 January, while the lower panels show the continuous monitoring results from the C925 HD system on 7 and 8 January. Gaps in the MTM time series correspond to periods where data were rejected due to adverse conditions, such as passing clouds that reduced stellar detections, or severe wind gusts that caused image streaking and temporary loss of tracking.

Despite these interruptions, the MTM-derived seeing values demonstrate high correlation with the DIMM measurements. As shown in Figure \ref{fig:seeing_evolution}, the MTM data points tightly cluster around the DIMM reference values across all three nights. The system accurately reproduces both the long-term evolution of atmospheric conditions and the short-term fluctuations. Furthermore, the nightly median seeing values derived from MTM are consistent with those reported by the DIMM, confirming the reliability of the MTM method for routine site monitoring.

\begin{figure*}
    \centering
    \begin{subfigure}{0.47\textwidth} 
    \includegraphics[width=\linewidth]{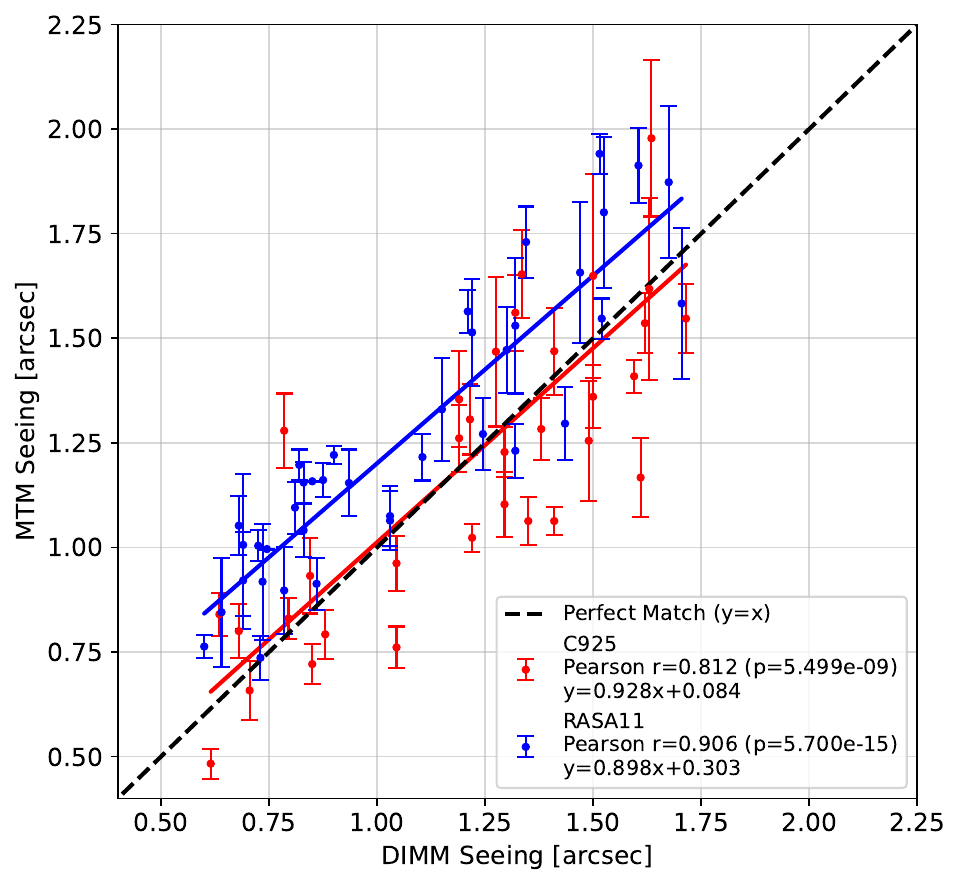} 
    \end{subfigure}
    \begin{subfigure}{0.47\textwidth}
    \includegraphics[width=\linewidth]{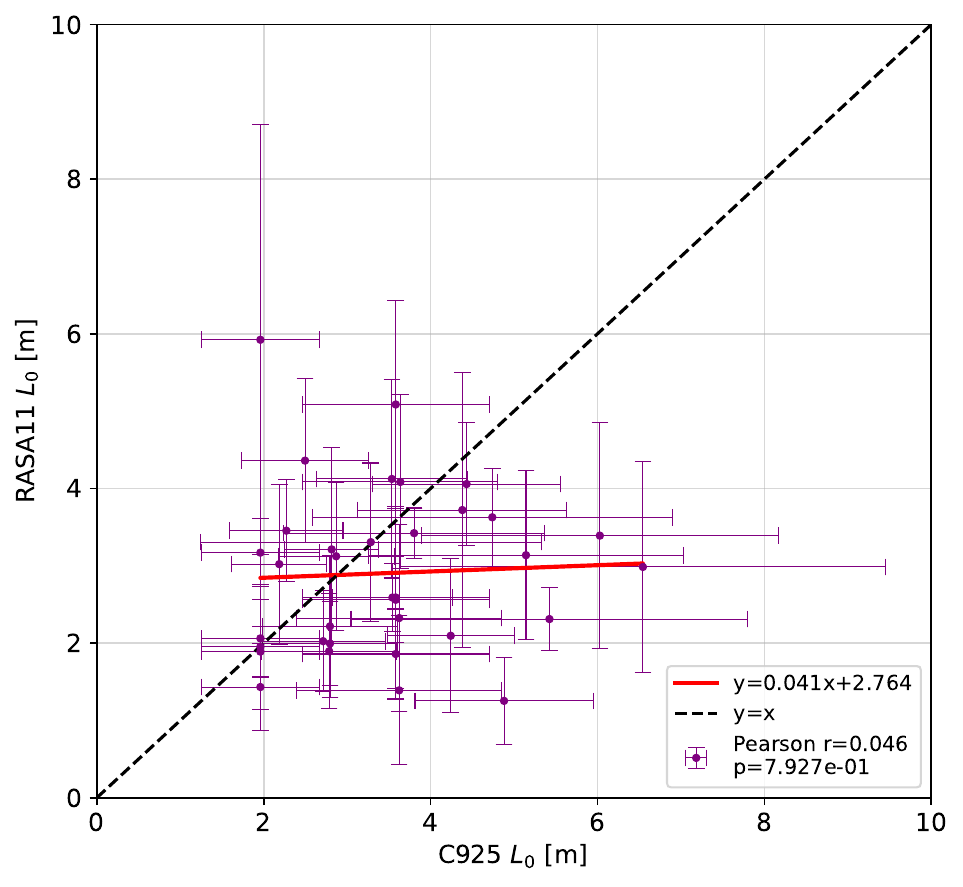}
    \end{subfigure}

    \caption{Correlation between MTM-derived seeing and simultaneous DIMM measurements obtained on 5 January. Left: MTM-derived seeing versus DIMM seeing. The red points represent data from the C925 HD configuration, while the blue points represent the RASA 11 configuration. Right: Comparison of the outer scale ($L_0$) measured by the RASA 11 and C925 HD systems, with extreme outliers excluded. Data points in both panels are plotted with their corresponding statistical uncertainties (error bars) derived from the inversion process. Solid lines indicate the linear regression fits, and the black dashed line represents the ideal $y=x$ relation.}
    \label{fig:seeing_correlation}
\end{figure*}

To quantitatively assess the measurement accuracy, Figure \ref{fig:seeing_correlation} presents the correlation between the MTM and DIMM seeing values obtained from the simultaneous observations on 5 January. Both configurations exhibit a strong linear correlation with the standard monitor. The C925 HD system achieves a Pearson correlation coefficient of $r = 0.812$ ($p\text{-value} < 0.001$) and a linear regression slope of 0.928. Similarly, the RASA 11 system shows a high correlation of $r = 0.906$ ($p\text{-value} < 0.001$) with a slope of 0.898.

It is worth noting that the regression slopes ($< 1$) indicate a damping of the measured turbulence strength variations compared to the DIMM. At high seeing values (strong turbulence), the MTM measurements tend to be lower than those of the DIMM. This underestimation is physically attributable to the finite exposure time ($t = 50$ ms), which causes temporal averaging of the wavefront tilt \citep{Hickson2019}, and the 10 km integration limit, which excludes the upper atmospheric contribution. Conversely, at low seeing values (weak turbulence), the MTM results may exhibit a slight overestimation due to the centroiding error. To rigorously quantify the measurement accuracy and assess the systematic offsets against the standard monitor, we computed the Pearson correlation coefficient ($r$), the mean bias ($B$), the bias-subtracted root-mean-square error (RMSE), and the mean ratio (MR) following standard instrument comparison methodologies \citep{Griffiths2024}. The statistical results for both MTM configurations relative to the DIMM are summarized in Table \ref{tab:comparison_metrics}.

\begin{table}
\centering
\caption{Statistical comparison between MTM configurations and the reference DIMM for the simultaneous seeing measurements on 5 January. The parameters include the Pearson correlation coefficient ($r$), bias-subtracted root-mean-square error (RMSE), mean bias ($B$), and mean ratio (MR).}
\label{tab:comparison_metrics}
\begin{tabular}{cccccc}
\hline
MTM & Reference  & $r$ & $B$  & RMSE  & MR \\
($Y_i$) & ($X_i$) & & (arcsec) & (arcsec) & \\
\hline
C925 HD & DIMM & 0.81 & $-0.02$ & 0.23 & 1.00 \\
RASA 11 & DIMM & 0.91 & 0.19 & 0.11 & 1.21 \\
RASA 11 & C925 HD & 0.79 & 0.10 & 0.18 & 1.12\\
\hline
\end{tabular}
\end{table}

Table \ref{tab:comparison_metrics} shows a noticeable positive bias of $0.19^{\prime\prime}$ for the RASA 11 configuration. This systematic positive bias is attributed to the specific detector used in the on-sky test. While the simulations assumed a monochrome camera, the actual field deployment for the RASA 11 system utilized a QHY268C color camera. The lower quantum efficiency (QE) of the color sensor resulted in a reduced SNR. This increased noise floor introduces a positive bias in the differential motion variance, which propagates into the inverted seeing values as a constant offset. In contrast, the C925 HD system utilized with a high-sensitivity monochrome camera (ASI1600MM) effectively minimized this error, resulting the bias to zero.

Despite these specific instrumental factors, the strong correlation between the MTM and DIMM seeing measurements, with Pearson correlation coefficients of $r=0.81$ and $r=0.91$ for the two instrumental configurations, provides robust observational validation of the method. Importantly, the fact that two substantially different optical systems—differing in aperture, focal length, and pixel scale—both reliably reproduced the seeing values demonstrates the instrument-independence of the technique. These results confirm that MTM offers a flexible and versatile solution for atmospheric turbulence monitoring, capable of delivering consistent performance across various hardware setups.

In addition to the integrated seeing, the outer scale of turbulence ($L_0$) is a critical parameter for optimizing large telescope instrumentation. The time series of the retrieved $L_0$ are presented on the secondary y-axis of Figure \ref{fig:seeing_evolution}, demonstrating the system's capability to monitor $L_0$ fluctuations. Furthermore, the right panel of Figure \ref{fig:seeing_correlation} presents a comparison of the $L_0$ measurements between the RASA 11 and C925 HD systems obtained on 5 January. The $L_0$ measurements exhibit noticeable scatter and do not strictly follow the ideal $y=x$ distribution. This deviation is primarily attributed to two factors. First, the retrieval of the outer scale is mathematically sensitive to the integrated seeing; thus, the slight systematic offset in the seeing measurements between the two systems inevitably propagates into the $L_0$ estimations. Second, although the two systems operated concurrently, their data acquisition was not strictly synchronized due to the significantly longer image readout time of the camera on the RASA 11 system. Despite these practical constraints, both systems successfully capture the macroscopic distribution range of the outer scale, predominantly clustering between 2 and 6 meters.

\subsection{Turbulence profile characteristics}

\begin{figure*} 
  \centering
  \begin{subfigure}{0.49\textwidth}
    \includegraphics[width=\linewidth]{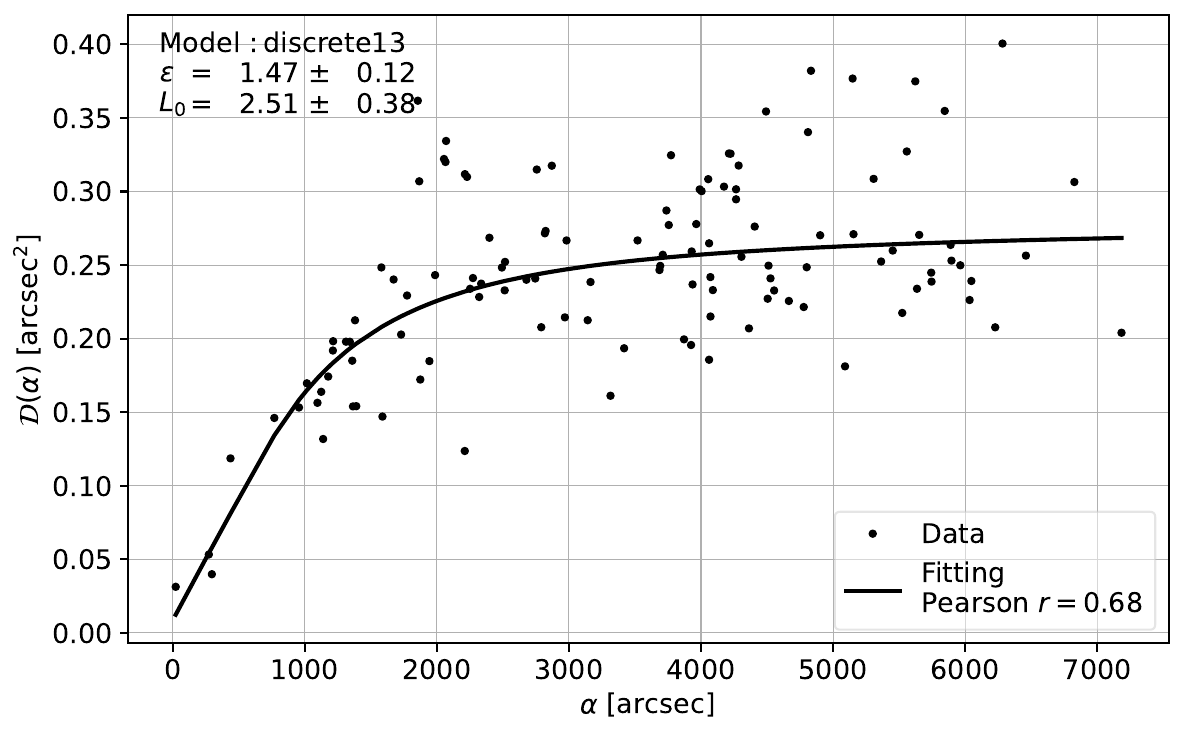}
  \end{subfigure}
  \begin{subfigure}{0.49\textwidth} 
    \includegraphics[width=\linewidth]{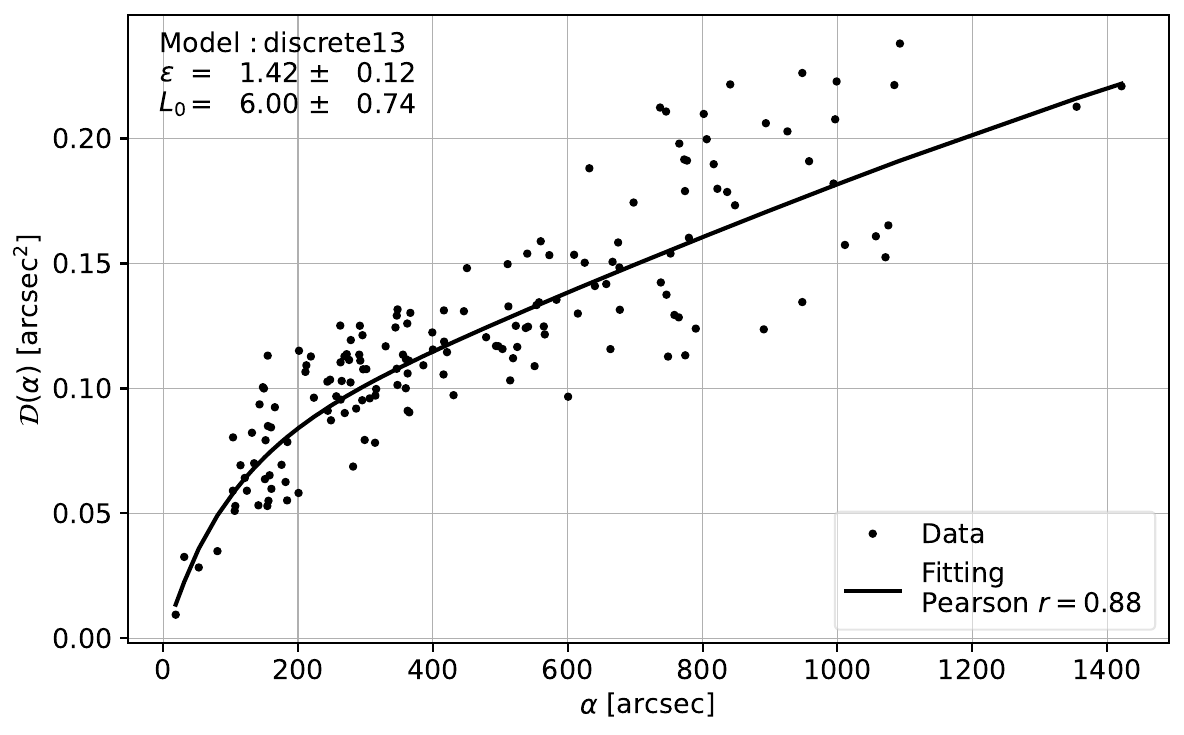} 
  \end{subfigure}
  
  \begin{subfigure}{0.36\textwidth}
    \includegraphics[width=\linewidth]{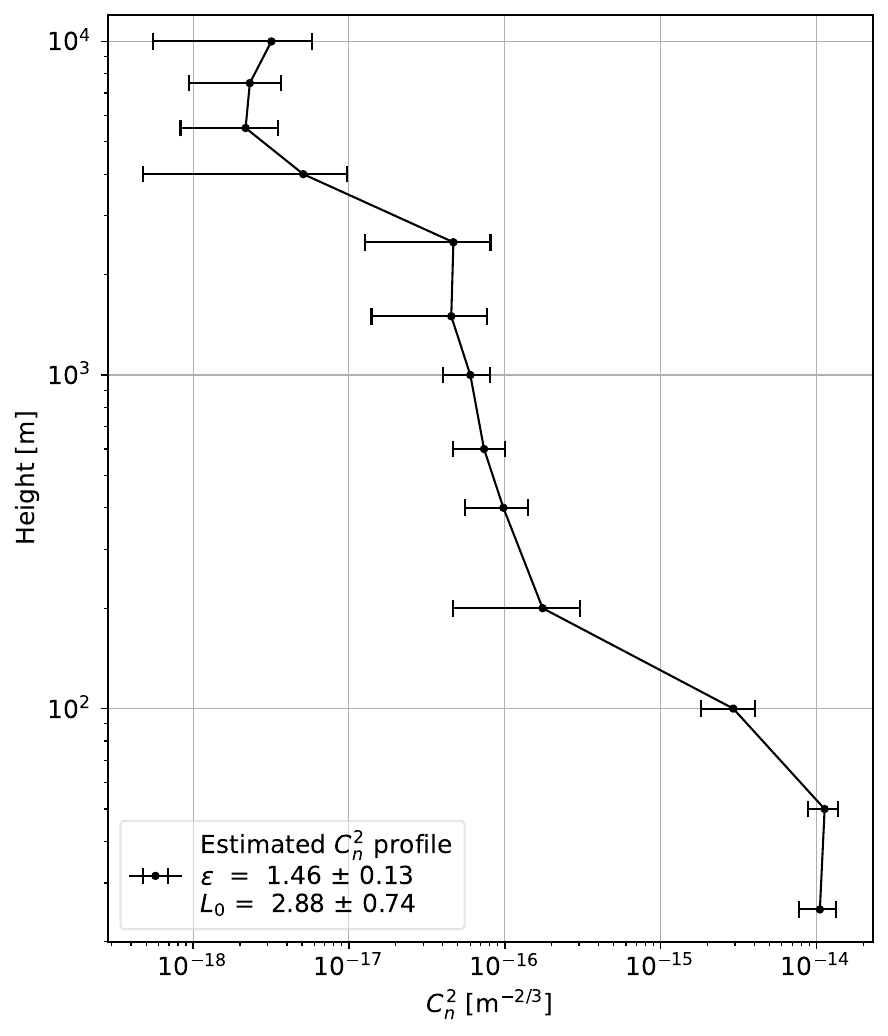}
  \end{subfigure}
  \begin{subfigure}{0.36\textwidth}
    \includegraphics[width=\linewidth]{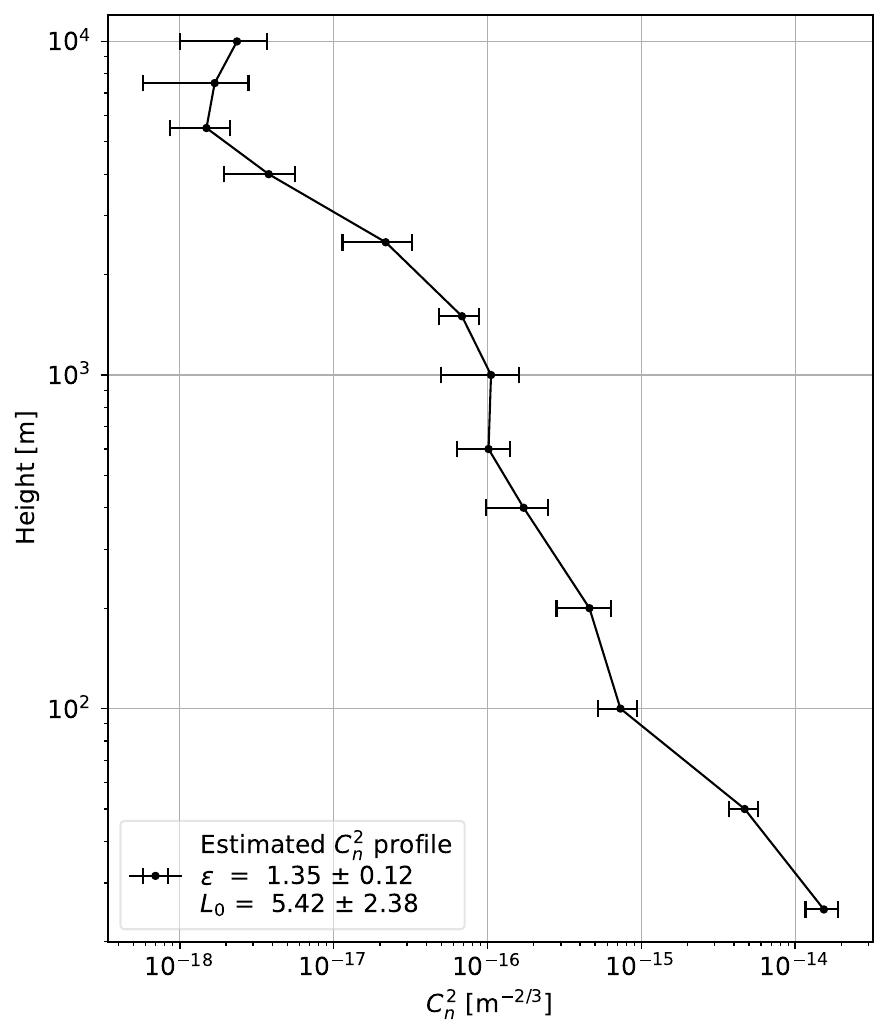}
  \end{subfigure}

  \caption{Snapshot of MTM inversion results obtained simultaneously from the RASA 11 (left) and C925 HD (right) systems on 5 January. Top panels: The observed differential image motion variance (dots) versus angular separation, overlaid with the best-fit model structure functions (solid lines). Bottom panels: Statistical results of the retrieved turbulence profiles from 30 MCMC realizations. Both configurations yield consistent seeing values and vertical profiles.}
  \label{fig:single_profile}
\end{figure*}

To examine the vertical structure of the turbulence, Figure \ref{fig:single_profile} presents a snapshot of the inversion results obtained simultaneously from the RASA 11 and C925 HD configurations targeting the open cluster M34 at 22:02 on 5 January. The upper panels display the observed differential image motion variance as a function of stellar separation $\alpha$. The lower panels show the corresponding vertical turbulence profiles derived from 30 MCMC realizations.

While both systems clearly reveal a stratified atmosphere with a dominant ground layer and a noticeable enhancement in turbulence strength around 10 km, a discrepancy is observed in the integrated seeing values. This offset of approximately $0.1''$ is consistent with the systematic bias identified in the correlation analysis (see Section \ref{section:seeing DIMM} or Table \ref{tab:comparison_metrics}). As previously discussed, this is primarily attributed to the lower signal-to-noise ratio of the RASA system's color camera. Despite this systematic offset in the ground layer, the consistent retrieval of high-height features validates the MTM's capability to resolve atmospheric stratification across different hardware setups.

\begin{figure*}
    \centering
    \includegraphics[width=1\linewidth]{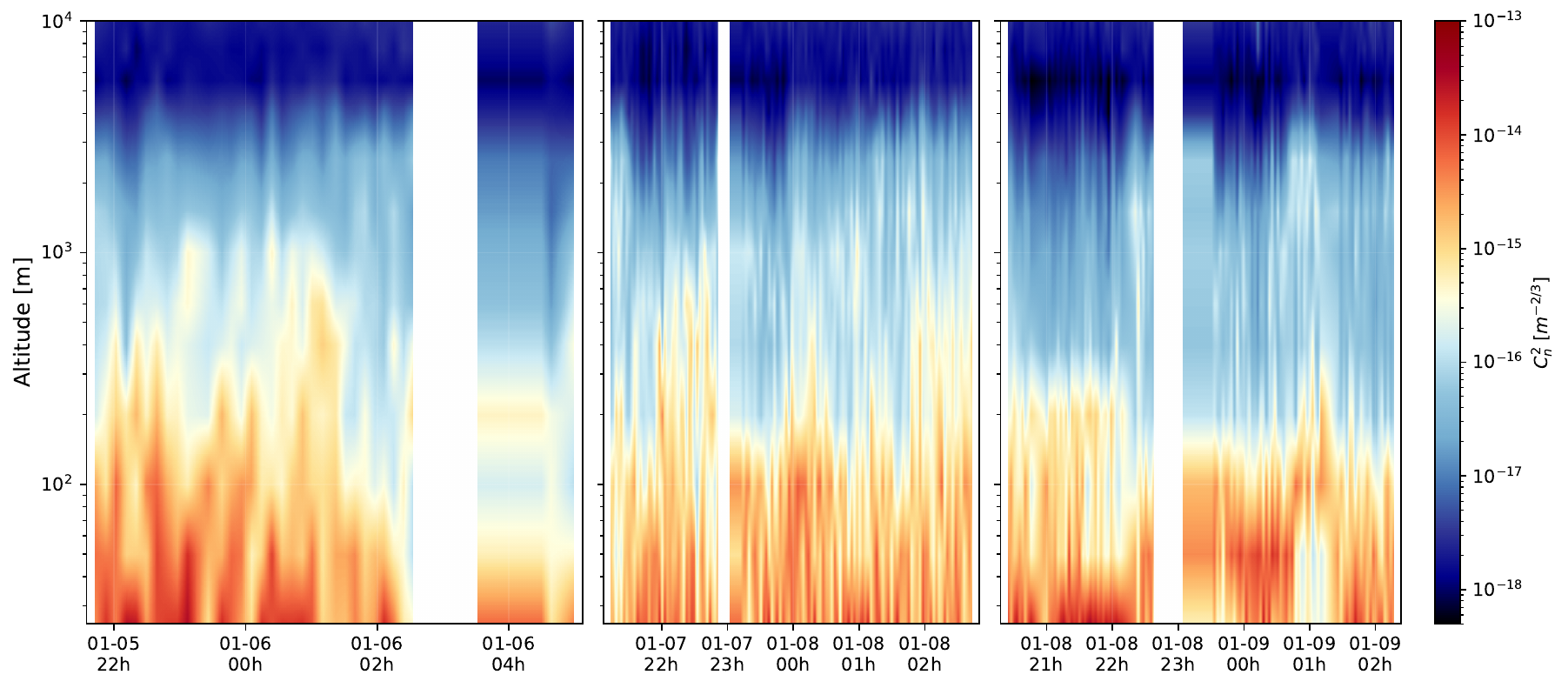}
    \caption{Spatiotemporal evolution of the refractive index structure constant $C_n^2$ measured over the three nights. The profiles are displayed as color-coded intensity maps, with height on a logarithmic scale. The color bar indicates the turbulence strength, ranging from weak turbulence (black) to strong optical turbulence (dark red). The atmosphere exhibits a pronounced stratification: a dominant, highly variable surface layer below 200 m, a stable free atmosphere reaching a minimum around 8 km, and a slight enhancement near the top of the profile.}
    \label{fig:stack_plot}
\end{figure*}

The temporal evolution of the turbulence stratification over the three nights is visualized in Figure \ref{fig:stack_plot}, which presents the time-series of $C_n^2$ profiles as a color-coded intensity map. This visualization effectively reveals the strong stratification and dynamic nature of the atmosphere above Daocheng. The vertical structure is clearly divided into distinct regimes. The optical turbulence is dominated by the surface layer (below $\sim 200$ m), indicated by the warm colors (red/orange) representing high $C_n^2$ values. This layer exhibits significant temporal intermittency, with rapid bursts of intense turbulence driving the fluctuations in the total seeing. Above the boundary layer, the $C_n^2$ values decrease significantly. The free atmosphere exhibits lower turbulence strength compared to the ground layer and appears predominantly calm (blue regions). Notably, the turbulence intensity reaches a minimum around 8 km, followed by a subtle re-intensification in the uppermost layers (above 8 km). Furthermore, the higher sampling cadence on 7 and 8 January (middle and right panels) reveals fine-grained vertical structures and rapid transient events that are smoothed out in the lower-cadence data of 5 January.

\begin{figure}
    \centering
    \includegraphics[width=1\linewidth]{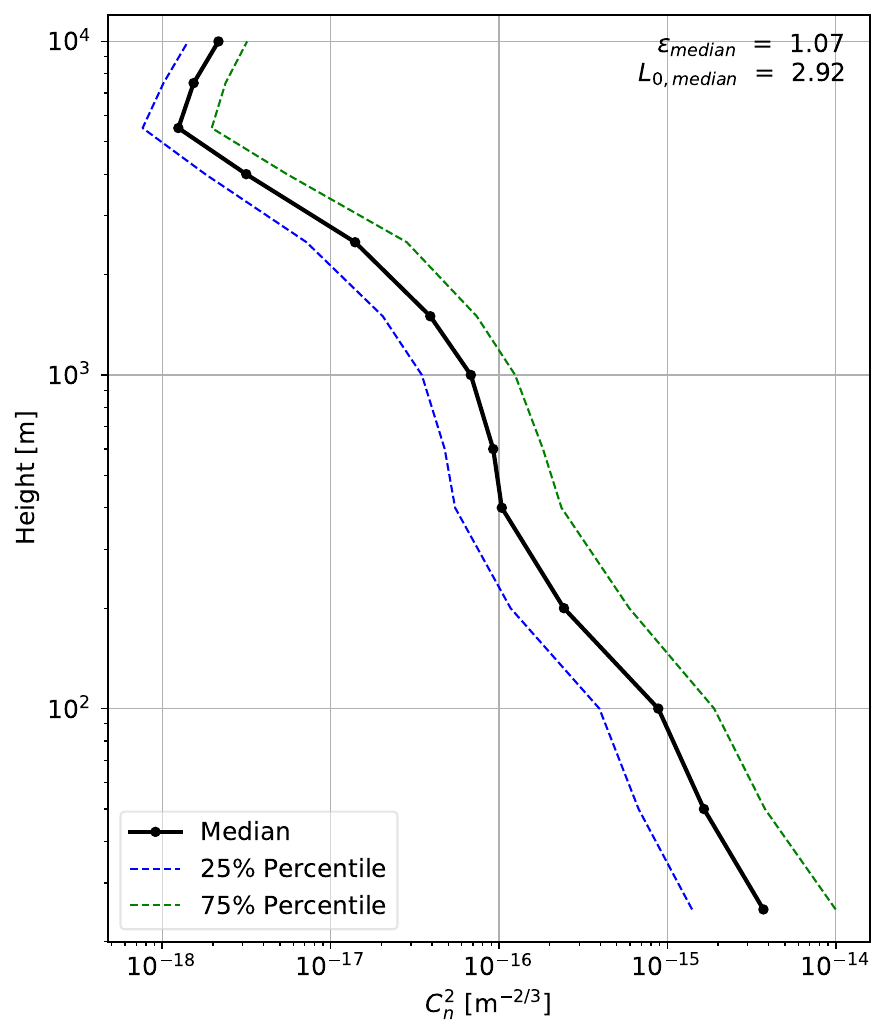}
    \caption{Statistical vertical turbulence profiles for the nights of 5, 7, and 8 January. The black lines with markers represent the median $C_n^2$ values at each height node, while the blue and green dashed lines indicate the 25th and 75th percentiles, respectively. A characteristic secondary peak around 10 km is visible, corresponding to the seasonal high-altitude wind shear.}
    \label{fig:median_profiles}
\end{figure}

Figure \ref{fig:median_profiles} summarizes the statistical vertical distribution of turbulence for the three observing nights. The solid lines represent the median $C_n^2$ profiles, while the dashed lines indicate the interquartile range (25th to 75th percentiles). The profiles exhibit a characteristic shape: a dominant ground layer that decreases rapidly with height, followed by a quiescent region in the middle troposphere. Interestingly, a distinct enhancement in turbulence strength is consistently observed in the height range of 8--10 km. This feature is likely associated with the tropopause and the subtropical jet stream. According to a recent statistical analysis of the Daocheng site using ERA5 meteorological reanalysis data \citep{Zhou2025}, the wind speed profiles in January typically exhibit a pronounced maximum near the 200\,hPa pressure level (approximately 10--12\,km altitude), corresponding to the seasonal position of the subtropical jet stream. The strong wind shear associated with this high-altitude jet generates optical turbulence, which is accurately captured by the MTM profiles as the observed increase in $C_n^2$ above 8 km. This agreement reinforces the physical reliability of the inversion results and illustrates the capability of the MTM technique to resolve subtle, meteorologically significant turbulence structures beyond the ground layer.

As discussed in Section \ref{sec:requirements}, the sensitivity to high-altitude turbulence depends on the response function saturation point, governed by the relation $\log(\alpha z / D) \approx 1.5$. In this study, we restricted the inversion upper height limit to 10 km to ensure a high density of samples at small separations, thereby guaranteeing the robust convergence of the inversion algorithm. The upper height limit of 10 km in our reconstructed profiles is a parameter choice for inversion stability, not an inherent limitation of this technique. Ensuring sufficient sampling of stellar pairs with angular separations within $\sim 100''$ would theoretically allow profiling up to $\sim 15$ km.  

\section{Conclusions}

We have presented a comprehensive investigation of the Multistar Turbulence Monitor (MTM), establishing it as a robust technique for retrieving vertical atmospheric turbulence profiles using compact, portable instrumentation. By combining theoretical analysis of response functions with numerical simulations and on-site observations, we systematically evaluated the method's performance and its dependence on instrumental parameters.

Our analysis of the instrumental requirements highlights the critical trade-off between field of view (sensitivity to ground layer) and pixel scale (centroiding precision). Numerical simulations based on the Hufnagel--Valley model demonstrate that the MTM inversion pipeline can robustly recover both the integrated seeing and the $C_n^2(z)$ profile, even under realistic noise conditions. Furthermore, extensive simulations across a broad range of atmospheric conditions (seeing from $0.6''$ to $1.8''$) confirm the linearity and stability of the inversion algorithm. These tests demonstrate that the MTM method can provide reliable turbulence estimates even with small, commercially available telescopes.

Three nights of observations at the Daocheng Astronomical Site provided rigorous validation of the method. We deployed two distinct configurations—a wide-field RASA 11 system and a high-resolution C925 HD system—operating simultaneously. Despite significant differences in aperture, focal length, and sampling, both systems provided consistent estimates of seeing and outer scale, as well as highly consistent reconstructed $C_n^2$ profiles. This consistency highlights the instrument-independence of the MTM technique and its suitability for deployment on a wide range of portable systems.

The turbulence profiles recovered at Daocheng show a prominent ground-layer component and a significant free-atmosphere contribution, in agreement with classical turbulence models. A secondary peak in the range of 8--10 km appears consistently in the reconstructions. This high-altitude feature aligns with the seasonal wind-speed maximum observed near the 200 hPa level, demonstrating the ability of MTM to detect subtle upper atmospheric structures driven by the jet stream. Furthermore, comparison with simultaneous DIMM measurements shows excellent agreement. The MTM-derived seeing closely follows the DIMM time series over the entire campaign, accurately reproducing both long-term evolution and short-timescale fluctuations.

The MTM shares certain conceptual similarities with other differential motion techniques, such as PDSL, but differs in several important respects. The MTM utilizes differential image motion between multiple stars, does not rely on a specific target such as the solar disc. These advantages make MTM a flexible and powerful technique for nighttime site testing and atmospheric characterization.

While the profiles presented here extend to 10 km, this does not represent the technique's limit. With optimized target selection to include more smaller separation pairs, the MTM technique is capable of resolving turbulence structures above 10 km.

Finally, while the current portable setups successfully demonstrated the observational feasibility of the MTM method, the use of a 50~ms exposure time introduced a wind smearing effect. Although strictly simultaneous high-resolution wind profiles were unavailable during our campaign, we can analytically estimate this temporal averaging effect. Utilizing the statistical January wind speed profile for the Daocheng site derived from ERA5 reanalysis \citep{Zhou2025}, coupled with our observed mean turbulence profile, we estimate an effective turbulence-weighted wind speed of approximately $9$~m/s. According to the analytical framework by \citet{Hickson2019}, this corresponds to a characteristic time scale $T_0 \approx 100$~ms. Our exposure time of 50~ms thus yields a ratio of $t/T_0 \approx 0.5$, which theoretically attenuates the measured differential structure function by approximately 5\%. This signal attenuation provides a physical explanation for why our measured seeing values exhibit a slight underestimation at the strong turbulence end, resulting in linear regression slopes (0.812 and 0.906) slightly less than unity relative to the DIMM. In future deployments, we plan to upgrade the system with higher-sensitivity scientific detectors to reduce the exposure time to 10~ms or less. This hardware improvement will inherently minimize these temporal averaging effects and enable a more precise characterization of rapid turbulence evolution under high wind-shear conditions.

In summary, the MTM method offers a versatile, low-cost, and effective solution for atmospheric turbulence profiling. Its demonstrated reliability across different hardware setups makes it particularly suitable for site testing and long-term monitoring at remote, high-altitude sites, supporting the site selection and optimization of future large-aperture astronomical facilities.

\section*{Acknowledgements}

This work is supported by the National Natural Science Foundation of China (NSFC) under grant nos. 12473089 and U2031148. PH acknowledges financial support from the Natural Sciences and Engineering Research Council of Canada, RGPIN-2019-04369.

\section*{Data Availability}
The data underlying this article will be shared on reasonable request to the corresponding author.



\bibliographystyle{mnras}
\bibliography{reference} 





\bsp	
\label{lastpage}
\end{document}